\documentclass[
prd,
nofootinbib,
preprintnumbers,
eqsecnum,
superscriptaddress,
showpacs
]{revtex4}

\usepackage{graphicx}
\usepackage{amsfonts}
\usepackage{amsmath}
\usepackage{amssymb}
\usepackage{amsxtra}
\usepackage{latexsym}
\usepackage{color}
\usepackage{colordvi}
\usepackage{xcolor}
\usepackage{epsfig}
\usepackage{array}
\usepackage{pifont}
\usepackage{braket}

\newcommand{\alinea}{\hspace*{\parindent}}

\newcommand{\re}{\text{Re}}
\newcommand{\tr}{\text{Tr}}

\def\krto{ {\,\,\lower .8ex\hbox {$\longrightarrow \atop k \rightarrow 0$}\,\,}}

\def\alinea{\hspace{\parindent}}
\def\bea{\begin{eqnarray} }
\def\beq{\begin{eqnarray} }

\def\eea{\end{eqnarray}}
\def\eeq{\end{eqnarray}}

\def\eq#1{Eq.~(\ref{#1})}

\begin{document} 

\title{Refining the detection of the zero crossing for the three-gluon vertex in symmetric and asymmetric momentum subtraction schemes}

\author{Ph.~Boucaud} 
\affiliation{ Laboratoire de Physique Th\'eorique (UMR8627), CNRS, Univ. Paris-Sud, Universit\'e Paris-Saclay, 91405 Orsay, France}
\author{F.~De Soto}
\affiliation{Dpto. Sistemas F\'isicos, Qu\'imicos y Naturales, 
Univ. Pablo de Olavide, 41013 Sevilla, Spain}
\author{J.~Rodr\'{\i}guez-Quintero}
\affiliation{Department of Integrated Sciences;  
University of Huelva, E-21071 Huelva; Spain.}
\affiliation{CAFPE, Universidad de Granada, E-18071 Granada, Spain}
\author{S.~Zafeiropoulos}
\affiliation{Department of Physics, College of William and Mary, Williamsburg, VA 23187-8795, USA}
\affiliation{Jefferson Laboratory, 12000 Jefferson Avenue, Newport News, VA 23606, USA}


\begin{abstract}

This article reports on the detailed study of the three-gluon vertex in four-dimensional $SU(3)$ Yang-Mills theory employing lattice simulations with large physical volumes and high statistics. A meticulous scrutiny of the so-called symmetric and asymmetric kinematical configurations is performed and it is shown that the associated form-factor changes sign at a given range of momenta. The lattice results are compared to the model independent predictions of Schwinger-Dyson equations and a very good agreement among the two is found.

\end{abstract}


\maketitle





\section{Introduction}
\alinea

The theory describing the strong interactions, Quantum Chromodynamics (QCD), is essentially a quantum field theory based on a local non-abelian gauge theory which, in its infrared (IR) sector, possesses a very rich  and intricate structure controlling its low-momentum dynamics. Phenomena such as confinement and chiral symmetry breaking, and hence the origin of most of the baryonic mass, clearly root in the IR sector of the theory. The understanding of this IR dynamics is crucial and, indeed, it has been very much boosted in the last few years, mainly due to very careful and detailed studies of the fundamental Green's functions of the theory in both lattice~\cite{Cucchieri:2006tf,Cucchieri:2008qm,Cucchieri:2007md,Cucchieri:2010xr,Bogolubsky:2009dc,Oliveira:2009eh,
Ayala:2012pb,Duarte:2016iko} and continuum QCD~\cite{Aguilar:2008xm,Boucaud:2008ky,Fischer:2008uz,RodriguezQuintero:2010wy,
Pennington:2011xs,Maris:2003vk,Aguilar:2004sw,Boucaud:2005ce,Fischer:2006ub,Kondo:2006ih,Binosi:2007pi,Binosi:2008qk,
Boucaud:2007hy,Dudal:2007cw,Dudal:2008sp,Kondo:2011ab,Szczepaniak:2001rg,Szczepaniak:2003ve,Epple:2007ut,
Szczepaniak:2010fe,Watson:2010cn,Watson:2011kv}. It is not worthless reminding here that those fundamental Green's functions are essential building blocks in the construction of any proper symmetry-preserving truncation of Schwinger-Dyson equations (SDEs) in order to define a tractable continuum bound-state problem able to reproduce the observable properties of hadrons\cite{Maris:2003vk,Chang:2009zb,Chang:2011vu,Qin:2011dd,Qin:2011xq,Bashir:2012fs,Eichmann:2012zz,
Cloet:2013jya,Binosi:2014aea}. 

The endeavors in obtaining two-point Green's function from large-volume lattice simulations~\cite{Cucchieri:2010xr,Bogolubsky:2009dc,Oliveira:2009eh,Ayala:2012pb,Duarte:2016iko} crystallized in our current well-known picture about the infrared finiteness of both the gluon propagator and ghost dressing function. This picture has been consistently interpreted by using a variety of different approaches,  such as the so-called refined Gribov-Zwanziger scenario~\cite{Dudal:2007cw,Dudal:2008sp}, effective descriptions based on the Curci-Ferrari model~\cite{Tissier:2010ts,Tissier:2011ey} or SDEs under different truncation schemes~\cite{Aguilar:2008xm,Boucaud:2008ky,Fischer:2008uz,RodriguezQuintero:2010wy,Pennington:2011xs}. In particular, one of the approaches within this last class~\cite{Aguilar:2008xm} is based on the combination of the pinch technique (PT) and the background field method (BFM). This latter PT-BFM framework~\cite{Binosi:2009qm} allows for a systematical rearranging of classes of diagrams in the nonperturbative expansion for the SDEs, leading to modified Green's functions obeying linear (abelian-like) Slavnov-Taylor identities (STI). A subtle realization of the Schwinger mechanism, within this framework, takes place and endows the gluon with an effective --dynamically acquired-- mass~\cite{Cornwall:1981zr,Bernard:1982my,Donoghue:1983fy,Philipsen:2001ip}. Furthermore, a profound connection emerges between the massless nature of the ghost propagator, the very deep IR behavior of the gluon and zero-momentum divergences of the three-gluon vertex that should be observed in some particular kinematic limits~\cite{Aguilar:2013vaa}. The latter is valid in the Landau gauge in three as well as four dimensions. The entanglement of these features stems precisely from how the mass-generation mechanism does remain transparent for the ghosts which, contrarily to the gluons, appear to flow in the quantum loops without the IR "{\it protection}" of a running mass. As a consequence of this, some of the form factors of the nonperturbative three-gluon vertex appear to be dominated in the IR by the nonperturbative ghost-loop contribution, taking negative values and diverging at vanishing momentum, precisely as a logarithm in four dimensions~\cite{Aguilar:2013vaa}. Hence, they change sign at a zero crossing which takes place in the IR. 

The same IR pattern has been also claimed by independent SDE analysis, employing various approaches and truncation schemes, in the three-gluon~\cite{Tissier:2011ey,Aguilar:2013vaa,Pelaez:2013cpa,Blum:2014gna,Eichmann:2014xya,Cyrol:2016tym} and four-gluon~\cite{Binosi:2014kka,Cyrol:2014kca} sectors; although the zero crossing is predicted to happen at such a low momentum that it is hard to be revealed by realistic lattice QCD simulations. Some studies in an $SU(2)$ lattice gauge theory had shown this expected pattern for the three-gluon vertex to emerge in three dimensions but failed to be conclusive in four~\cite{Cucchieri:2006tf,Cucchieri:2008qm}; and, very recently, two independent investigations~\cite{Athenodorou:2016oyh,Duarte:2016ieu} in $SU(3)$ have provided with a preliminary confirmation for this to happen for the three-gluon vertex also in four dimensions. 

In this article, we aimed at the completion of the preliminary work presented in~\cite{Athenodorou:2016oyh} by largely increasing the statistics --the number of gauge-field configurations as well as the number of different lattice set-ups-- and by performing a more elaborated and refined analysis, based on the SDE interpretation of the IR three-gluon behavior, intended to a better detection of the zero crossing and to solidify our theoretical understanding of its happening. To this purpose, we have performed $SU(3)$ simulations in large four-dimensional volumes with two lattice actions (Wilson-plaquette and tree-level Symanzik), computed the three-gluon Green's function and projected out the relevant form factor in two different kinematical configurations.  We introduce all the definitions and carefully describe the procedure for the extraction of the form factors from the nonperturbative lattice Green's functions, the renormalization scheme and the connection with the running strong coupling in section \ref{sec:3g}. The section \ref{sec:Lat} is devoted to provide with the details of the simulations, the lattice actions and set-ups, and to display the results for the running strong coupling and the form factors. The analysis of the lattice results can be found in section \ref{sec:SDE}, where we also discuss their interpretation; and, finally, we conclude in section \ref{sec:conclusions}.


\section{The three-gluon Green's function and the running coupling}
\label{sec:3g}

\subsection{Definitions of the connected and 1-PI vertex functions}

Let us first define the connected three-gluon vertex as the correlation function of the following gauge fields, $\widetilde{A}$, taken in Fourier space at the three momenta $p,q$ and $r$, such that $p+q+r=0$,

\begin{eqnarray}\label{eq:3gdef}
{\cal G}^{abc}_{\alpha\mu\nu}\left(p,q,r\right) 
=  \ \langle \widetilde{A}_{\mu}^{a}(p) \widetilde{A}_{\nu}^{b}(q) \widetilde{A}_{\rho}^{c}(q) \rangle \ 
= f^{abc} \ {\cal G}_{\alpha \mu \nu}\left(p,q,r\right)
\end{eqnarray}

\noindent
where the sub (super) indices are Lorentz (color) ones and the average $\langle \cdot \rangle$ indicates functional integration over the gauge space. For the sake of convenience, we will work in the Landau gauge, in which only the projection onto the totally antisymmetric color tensor, $f^{abc}$, survives for the three-gluon Green's function, making \eq{eq:3gdef} plainly general. In terms of the usual 1-particle irreducible (1-PI) vertex function, $\Gamma$, and the gluon propagator 
\begin{eqnarray}\label{eq:prop}
\Delta^{ab}_{\mu\nu}\left(p\right) \ = \ \langle \widetilde{A}^a_\mu(p) \widetilde{A}^b_\nu(-p) \rangle \ = \ 
\Delta(p^2) \delta^{ab}  P_{\mu \nu}(p) \ ,
\end{eqnarray}
with $P_{\mu\nu}(p) = \delta_{\mu\nu}-p_\mu p_\nu/p^2$, the three-gluon Green's function can be recast as
\begin{eqnarray}\label{eq:3gdef2}
{\cal G}^{abc}_{\alpha \mu \nu}\left(p,q,r\right)  &=& g f^{a'b'c'} 
\Gamma_{\alpha' \mu' \nu'}\left(p,q,r\right)  \ \Delta^{a'a}_{\alpha' \alpha}(p) \Delta^{b'b}_{\mu' \mu}(q) \ \Delta^{c'c}_{\nu' \nu}(r) \\ \label{eq:3gdef3}
&=& g f^{abc} \Gamma_{\alpha' \mu' \nu'}\left(p,q,r\right) \Delta(p^2) \Delta(q^2) \Delta(r^2) P_{\alpha' \alpha}(p) 
P_{\mu' \mu}(q) P_{\nu' \nu}(r) \ ;
\end{eqnarray}
where $g$ is the strong coupling constant. Thus, the vertex function ${\cal G}_{\alpha \mu \nu}$, introduced in \eq{eq:3gdef}, can be easily seen in \eq{eq:3gdef3} to read in terms of gluon dressing functions and transversal projectors such that the transversality condition 
\begin{equation}\label{eq:Gtrans}
p_\alpha {\cal G}_{\alpha \mu \nu}(p,q,r) = q_\mu {\cal G}_{\alpha \mu \nu}(p,q,r)=r_\nu {\cal G}_{\alpha \mu \nu}(p,q,r)=0
\end{equation} 
is made clearly apparent. A basis of four tensors can generally describe the subspace where the vertex function ${\cal G}_{\alpha \mu \nu}$ is embedded\cite{Ball:1980ax,Aguilar:2013vaa,Eichmann:2014xya}. However, in what follows, we will specialize in two particular momenta configurations for the three-gluon vertex which, from now on, will be called `{\it symmetric}' and `{\it asymmetric}'. The first one corresponds to the case defined by $p^2=q^2=r^2$ and $p\cdot q=p\cdot r=q\cdot r = -p^2/2$, in which the subspace of the totally transverse tensors, observing \eq{eq:Gtrans}, has dimension two. Hence a basis is made by only two tensors, namely
\begin{eqnarray}\label{eq:basis}
\lambda^{\rm tree}_{\alpha \mu\nu}(p,q,r) &=&  \Gamma^{(0)}_{\alpha' \mu'\nu'}(p,q,r) P_{\alpha'\alpha}(p) P_{\mu'\mu}(q) P_{\nu'\nu}(r) \ , \nonumber \\
\lambda^{\rm S}_{\alpha\mu\nu}(p,q,r) &=& \frac{(p-q)_{\nu} (q-r)_{\alpha} (r-p)_{\mu}}{p^2} \ ,
\end{eqnarray}
where $\Gamma^{(0)}_{\alpha \mu \nu}(p,q,r) = \delta_{\alpha\mu} (p-q)_{\nu} +  \delta_{\mu\nu} (q-r)_{\alpha} 
+ \delta_{\nu\alpha} (r-p)_{\mu}$ stands for the perturbative tree-level tensor of the three-gluon vertex; and one can write
\begin{eqnarray}\label{eq:calG}
{\cal G}_{\alpha\mu\nu}(p,q,r) \ = \ T^{\rm sym}(p^2) \ \lambda^{\rm tree}_{\alpha \mu\nu}(p,q,r) \ + \ 
S^{\rm sym}(p^2) \ \lambda^{\rm S}_{\alpha \mu\nu}(p,q,r) \ .
\end{eqnarray}
Now, taking advantage of that the transverse projector acts over the subspace defined by the basis \eqref{eq:basis} as the identity, 
\begin{eqnarray}
\lambda^{\rm tree}_{\alpha' \mu'\nu'}(p,q,r) P_{\alpha'\alpha}(p) P_{\mu'\mu}(q) P_{\nu'\nu}(r) & = &  \lambda^{\rm tree}_{\alpha \mu\nu}(p,q,r)  \nonumber \\ 
\lambda^{\rm S}_{\alpha'\mu'\nu'}(p,q,r) P_{\alpha'\alpha}(p) P_{\mu'\mu}(q) P_{\nu'\nu}(r) &=& \lambda^{\rm S}_{\alpha\mu\nu}(p,q,r) \ , 
\end{eqnarray}
\eq{eq:calG} can be rewritten as 
\begin{eqnarray}
{\cal G}_{\alpha\mu\nu}(p,q,r) &=& \left( T^{\rm sym}(p^2)  \lambda^{\rm tree}_{\alpha' \mu'\nu'}(p,q,r) +  
S^{\rm sym}(p^2)  \lambda^{\rm S}_{\alpha' \mu'\nu'}(p,q,r) \right)  
\nonumber \\ 
& & P_{\alpha'\alpha}(p) P_{\mu'\mu}(q) P_{\nu'\nu}(r) \ ,
\end{eqnarray}
that, compared to \eq{eq:3gdef3}, leads to the following decomposition for the transverse 1-PI vertex function
\begin{eqnarray}
g \Gamma_{\alpha\mu\nu}(p,q,r)  &=& 
\frac{ T^{\rm sym}(p^2)}{\Delta^3(p^2)} \ \lambda^{\rm tree}_{\alpha\mu\nu}(p,q,r) 
\ + \  \frac{ S^{\rm sym}(p^2)}{\Delta^3(p^2)} \ \lambda^{\rm S}_{\alpha\mu\nu}(p,q,r) 
\\
&=& g \Gamma_T^{\rm sym}(p^2) \ \lambda^{\rm tree}_{\alpha\mu\nu}(p,q,r) 
\ + \  g \Gamma_S^{\rm sym}(p^2) \ \lambda^{\rm S}_{\alpha\mu\nu}(p,q,r) \ .
\label{eq:Gammadecomp}
\end{eqnarray}
Thus, focusing on the form factor for the tree-level tensor, we get
\begin{equation}\label{eq:1pi}
 T^{\rm sym}(p^2) = g \Gamma_T^{\rm sym}(p^2) \Delta^3(p^2) \ ,
\end{equation}
where, in particular, the $T^{\rm sym}(p^2)$ form factor can be projected out through
\begin{eqnarray}\label{eq:projsym}
 T^{\rm sym}(p^2) = \frac{{\cal G}_{\alpha\mu\nu}(p,q,r) {\cal W}_{\alpha\mu\nu}(p,q,r)}{{\cal W}_{\alpha\mu\nu}(p,q,r) {\cal W}_{\alpha\mu\nu}(p,q,r)}
\end{eqnarray}
with 
\begin{eqnarray}
{\cal W}_{\alpha\mu\nu}(p,q,r) \ = \  \lambda^{\rm tree}_{\alpha\mu\nu}(p,q,r) + \frac{1}{2} \lambda^{\rm S}_{\alpha\mu\nu}(p,q,r) \ .
\end{eqnarray}

The second special momenta configuration that we will focus our attention on, is the so-called `{\it asymmetric}' one, and is defined by taking the $p \to 0$ limit, while keeping at the same time the condition $q^2=r^2=-q\cdot r$. In this case, in the Landau gauge, only one transverse tensor can be constructed and it appears to be the $p\to 0$ limit of the tree-level one
\bea
\lambda^{\rm tree}_{\alpha\mu\nu}(0,q,-q) \ = \ \Gamma^{(0)}_{\alpha\mu'\nu'}(0,q,-q) P_{\mu'\mu}(q) P_{\nu'\nu}(q)
\ = \  2  \ q_{\alpha} P_{\mu\nu}(q)  \ .
\eea
It is straightforward to notice that the second tensor in \eq{eq:basis}, $\lambda^{\rm S}_{\alpha\mu\nu}$, is only transverse if the `{\it symmetric}' condition $p^2=q^2=r^2$ is met. This is clearly not the case when the `{\it asymmetric}' configuration of momenta is considered and, indeed, the $p\to 0$ limit of this second tensor would result in a totally longitudinal structure, $\lambda^{\rm S}_{\alpha\mu\nu} \sim q_\alpha q_\mu q_\nu$.

Thus, we are left with a single form factor that can be projected out through 
\begin{eqnarray}\label{eq:projasym}
 T^{\rm asym}(q^2) = \frac{{\cal G}_{\alpha\mu\nu}(0,q,-q) {\cal \widetilde{W}}_{\alpha\mu\nu}(0,q,-q)}{{\cal \widetilde{W}}_{\alpha\mu\nu}(0,q,-q) {\cal \widetilde{W}}_{\alpha\mu\nu}(0,q,-q)}
\end{eqnarray}
where now
\begin{eqnarray}
{\cal \widetilde{W}}_{\alpha\mu\nu}(0,q,-q) \ = \  \lambda^{\rm tree}_{\alpha\mu\nu}(0,q,-q) \ .
\end{eqnarray}
This form factor relates with the 1-PI vertex function as follows
\begin{equation}\label{eq:1piasym}
T^{\rm asym}(q^2) \ = \ g \Gamma^{\rm asym}_T(q^2) \Delta(0) \Delta^2(q^2) \ ,
\end{equation}
where the $p\to 0$ limit brings here a dressing function evaluated at vanishing momentum that, as will be seen below, appears to be the source of additional statistical noise in extracting a nonperturbative signal for the form factor {\it via} \eq{eq:projasym} from lattice QCD.  

\subsection{The $R$-projector}

Aiming at a direct comparison with the lattice estimates for the three-gluon vertex given in refs.~\cite{Cucchieri:2006tf,Cucchieri:2008qm}, the authors of ref.~\cite{Aguilar:2013vaa} define a particular quantity projected out through the so-called $R$-projector (constructed with the tree-level tensor, $\Gamma^{(0)}$), 
\begin{eqnarray} \label{eq:RLatt}
R(p^2) &=& 
 \frac{g \Gamma^{(0)}_{\alpha\mu\nu}(p,q,r)  {\cal G}_{\alpha\mu\nu}(p,q,r)}{g \Gamma^{(0)}_{\alpha\mu\nu}(p,q,r) P^{\alpha\alpha'}(p)  \Delta(p^2) P^{\mu\mu'}(q) \Delta(q^2) \Delta(r^2) P^{\nu\nu'}(r) g \Gamma^{(0)}_{\alpha'\mu'\nu'}(p,q,r)} 
 \\
&=& \frac{\Gamma^{(0)}_{\alpha\mu\nu}(p,q,r) P^{\alpha\alpha'}(p)  P^{\mu\mu'}(q)  P^{\nu\nu'}(r) \Gamma_{\alpha'\mu'\nu'}(p,q,r)}{\Gamma^{(0)}_{\alpha\mu\nu}(p,q,r) P^{\alpha\alpha'}(p)  P^{\mu\mu'}(q)  P^{\nu\nu'}(r) \Gamma^{(0)}_{\alpha'\mu'\nu'}(p,q,r)} \ ,
\label{eq:Rsde}
\end{eqnarray}

\noindent where \eq{eq:RLatt} corresponds to the quantity evaluated in ref.~\cite{Cucchieri:2006tf,Cucchieri:2008qm} from lattice QCD (see Eq.(20) of~\cite{Cucchieri:2006tf}), adapted to our notation here; which is rewritten in~\cite{Aguilar:2013vaa} as it reads in \eq{eq:Rsde}, in terms of the full nonperturbative vertex functions, $\Gamma_{\alpha\mu\nu}$,  although explicitly brought to transversity by the appropriate projectors. 

Now, for the sake of completion, we will apply Eqs.~(\ref{eq:basis}) and (\ref{eq:Gammadecomp}) in \eq{eq:Rsde} and then, 
for the symmetric momenta configuration, get
\begin{eqnarray}\label{eq:R}
R(p^2) &=& \frac{{\lambda^{\rm tree}}^{\alpha\mu\nu}(p,q,r) \ \left[ \Gamma^{\rm sym}_T(p^2) \ \lambda^{\rm tree}_{\alpha\mu\nu}(p,q,r) + \Gamma^{\rm sym}_S(p^2) \ \lambda^{\rm S}_{\alpha\mu\nu}(p,q,r) \right]}
{{\lambda^{\rm tree}}^{\alpha\mu\nu}(p,q,r) \ \lambda^{\rm tree}_{\alpha\mu\nu}(p,q,r)} \nonumber \\
&=& \Gamma^{\rm sym}_T(p^2) \ - \ \frac{6}{11} \ \Gamma^{\rm sym}_S(p^2)  \ ,
\end{eqnarray}
which shows that the $R$-projector applied to the three-gluon Green's function results in a combination of the two form factors for the tensors in \eq{eq:basis}, which form a basis for the dimension-two transverse subspace in the symmetric case.  
 
On the other hand, in the case of the asymmetric configuration of momenta, as the transverse subspace is only of dimension-one, the action of the $R$-projector, 
\begin{eqnarray}\label{eq:Gproasym}
\widetilde{R}(q^2) \ = \ \frac{\Gamma^{(0)}_{\alpha\mu\nu}(0,q,-q) P^{\mu\mu'}(q)   P^{\nu\nu'}(q) \Gamma_{\alpha\mu'\nu'}(0,q,-q)}{\Gamma^{(0)}_{\alpha\mu\nu}(0,q,-q) P^{\mu\mu'}(q)   P^{\nu\nu'}(q)  \Gamma^{(0)}_{\alpha\mu'\nu'}(0,q,-q)} 
\ = \ \Gamma^{\rm asym}_T(q^2) \ ,
\end{eqnarray}
is equivalent to \eq{eq:projasym}.

\subsection{Renormalization and the running coupling}
\label{subsec:ren}

All the quantities that have been introduced so far are bare and one should everywhere understand an implicit dependence on the regularization cut-off. We apply now a given renormalization procedure where any renormalized Green's functions ought to be understood as the correlation function of renormalized gauge fields, $\widetilde{A}_R = Z_3^{-1/2}\widetilde{A}$; such that

\begin{eqnarray} \label{eq:renprop0}
\Delta_R(p^2;\mu^2) &=& Z_3^{-1}(\mu^2) \Delta(p^2) \ ,  \\
T_R(p^2;\mu^2) &=& Z_3^{-3/2}(\mu^2) T(p^2) \ , 
\label{eq:renT}
\end{eqnarray}

\noindent 
where \eq{eq:renT} is formally equivalent for both the asymmetric and symmetric cases, $\mu^2$ being the subtraction momentum. Then, specifying to the so-called MOM renormalization schemes, which are defined by imposing that all Green's functions take their tree-level value at the subtraction point for a particular choice of momenta configuration, one has for symmetric case

\begin{eqnarray}
\Delta_R(p^2;p^2) &=&  Z_3^{-1}(p^2) \Delta(p^2) \ = \frac 1 {p^2} \ , \label{eq:renprop} \\
T^{\rm sym}_R(p^2;p^2) &=&  Z_3^{-3/2}(p^2) T^{\rm sym}(p^2) \ = \ \frac{g^{\rm sym}_R(p^2)}{p^6}  \label{eq:ren3g}\ , 
\end{eqnarray}
and 
\begin{eqnarray}
T^{\rm asym}_R(p^2;p^2) &=&  Z_3^{-3/2}(p^2) T^{\rm asym}(p^2) \ = \ \Delta_R(0;p^2) \ \frac{g^{\rm asym}_R(p^2)}{p^4}  \label{eq:ren3gasym}\ , 
\end{eqnarray}
for the asymmetric. 

Then, \eq{eq:renprop} yields the renormalization constant $Z_3$ as a function of the bare propagator, while \eq{eq:ren3g} and \eq{eq:ren3gasym} provide us with the running coupling defined, respectively, in symmetric and asymmetric MOM schemes, 
\begin{eqnarray}\label{eq:g3g}
g_R^{\rm sym}(p^2) &=& p^3 \frac{T^{\rm sym}(p^2)}
{\left[\Delta(p^2)\right]^{3/2}} \ = \ p^3 \frac{T^{\rm sym}_R(p^2,\mu^2)}
{\left[\Delta_R(p^2,\mu^2)\right]^{3/2}} \ , \\
g_R^{\rm asym}(p^2) &=& p^3 \frac{T^{\rm asym}(p^2)}{\left[\Delta(p^2)\right]^{1/2} \Delta(0)} 
\ = \ p^3 \frac{T^{\rm asym}_R(p^2;\mu^2)}{\left[\Delta_R(p^2;\mu^2)\right]^{1/2} \Delta_R(0;\mu^2)} \ .
\label{eq:g3gasym}
\end{eqnarray}
Both are nonperturbative definitions for the QCD running coupling, that have been extensively studied on the lattice~\cite{Alles:1996ka,Boucaud:1998bq,Boucaud:2000nd,Boucaud:2001st,DeSoto:2001qx,Boucaud:2013jwa}. Other MOM schemes based on different 
QCD vertices and kinematical configurations, as that for the ghost-gluon vertex~\cite{vonSmekal:1997ohs,Sternbeck:2007br,Boucaud:2008gn,vonSmekal:2009ae}, lead to alternative nonperturbative definitions although they can be related at any order in perturbation 
theory~\cite{Chetyrkin:2000fd,Chetyrkin:2000dq}. Here, in obtaining \eq{eq:g3g} and \eq{eq:g3gasym}, we use first \eq{eq:ren3g} and \eq{eq:ren3gasym}, respectively, in order to express the running coupling in terms of bare Green's functions (replacing $Z_3$ by its \eq{eq:renprop} bare result); and only then, after realizing that the bare quantities appear as a renormalization-group independent (RGI) combination, as they should, replace them by their renormalized counterparts in the rightmost hand-sides. 

In what concerns the 1-PI vertex functions,  after applying the renormalization prescription to \eq{eq:1pi} and \eq{eq:1piasym}, respectively, for symmetric and asymmetric cases, one is left with
\begin{eqnarray}\label{eq:GammaR}
T_R^{\rm sym}(p^2;\mu^2) = g^{\rm sym}_R(\mu^2) \Gamma_{T,R}^{\rm sym}(p^2;\mu^2) \Delta_R^3(p^2;\mu^2) \ , \\
T_R^{\rm asym}(p^2;\mu^2) \ = \ g^{\rm asym}_R(\mu^2) \Gamma_{T,R}^{\rm asym}(p^2;\mu^2) \Delta_R(0;\mu^2) \Delta_R^2(p^2;\mu^2) \ ; \label{eq:GammaRasym}
\end{eqnarray}
where $\Gamma_{T,R}^{i}(p^2,p^2)=1$, for $i$ indicating either the symmetric (sym) or the asymmetric (asym) momenta configuration. Then, from either \eq{eq:GammaR} and \eq{eq:g3g} or \eq{eq:GammaRasym} and \eq{eq:g3gasym},  the renormalized 1-PI vertex functions can be calculated for, respectively, the symmetric and asymmetric momenta configuration and read in both cases
\begin{eqnarray}\label{eq:gGammaR}
g^i_R(\mu^2) \ \Gamma^i_{T,R}(p^2;\mu^2) = \frac{g^i_R(p^2)}{\left[p^2\Delta(p^2;\mu^2)\right]^{3/2} } \ .  
\end{eqnarray}
This last result is of special interest because it establishes a connection between the three-gluon MOM running coupling\footnote{Indeed, the quantity that can be generally found in literature is $\alpha(\mu^2)=g^2(\mu^2)/(4\pi)$ which, by squaring the signal of $g$ obtained from the lattice misses the existence of a change of sign, and hence a zero-crossing, at very deep IR momentum, for the vertex itself.}, which many lattice and continuum studies have paid attention to, and the vertex function of the amputated three-gluon Green's function, a relevant ingredient within the tower of (truncated) SDEs conceived to address the nonperturbative dynamics of QCD. Both quantities appear related only by the dimensionless gluon propagator dressing function, $D(p^2)=p^2 \Delta(p^2)$, which, after the intensive activity developed during the past decade~\cite{Aguilar:2004sw,Boucaud:2005ce,Aguilar:2006gr,Boucaud:2007hy,Dudal:2007cw,Aguilar:2008xm,Boucaud:2008ky,Dudal:2008sp,Fischer:2008uz,Aguilar:2009nf,Fischer:2009tn,Aguilar:2010gm}, is very well understood and accurately known.

\section{Lattice QCD results}
\label{sec:Lat}

Let us start this section with a reminder of how the three-gluon Green's function is computed from a lattice field theory simulation. In the purpose of concluding about qualitative features for its deep infrared behavior, one can simulate lattice volumes in physical units as large as possible but work in the quenched approximation, under the working assumption that light dynamical quarks will only affect quantitatively this behavior. The necessity to focus on the quenched (pure gauge) theory is dictated by the hefty numerical cost imposed by a very large volume simulation with dynamical fermions. To be more specific one needs to consider that the most prominent collaborations employ ensembles of gauge fields whose physical volumes are usually around 3 fm which is three times smaller than the smallest physical volume that exhibits data points in the deep IR regime where the zero crossing takes place. Sensu stricto, we are concluding here about the zero-crossing of the three-gluon Green's function in pure Yang-Mills theory. All the other recent lattice $SU(3)$ investigations~\cite{Athenodorou:2016oyh,Duarte:2016ieu} as well as the previous $SU(2)$ lattice studies~\cite{Cucchieri:2006tf,Cucchieri:2008qm} also work under the quenched approximation and concluded about a theory without light quarks. It cannot be then excluded that the effect of the light quarks might wipe out the zero-crossing, as it happens to take place in the far infrared. However, the motivation for the current and past studies in the pure Yang-Mills QCD is two-sided. (i) The zero-crossing has been claimed to happen as a purely gluodynamical effect in refs.~\cite{Tissier:2011ey,Aguilar:2013vaa,Pelaez:2013cpa,Blum:2014gna,Eichmann:2014xya,Cyrol:2016tym}, by using the SDE formalism, and then argued to shift down to deeper momenta by the presence of light quarks~\cite{Williams:2015cvx}. The study of that effect in pure gluodynamics deserves interest {\it per se} but, if the latter is true, can also provide with noteworthy qualitative information about the IR behavior of the QCD gluonic Green's functions. (ii) The mechanism we invoked here to explain the three-gluon sign changing and its zero-crossing, as will be discussed in the next section, is the same one, related to the nonperturbative ghost loop diagram contributing to the gluon self-energy, producing a peak for the gluon propagator at a deep IR non-zero momentum. This gluon propagator peak and the mechanism behind it have been discussed in ref.~\cite{Binosi:2016xxu} within the context of the restoration of chiral symmetry in QCD with increasing number of light flavors. Thus, although in pure gluodynamics, our analysis relies on a mechanism with, presumably, profound implications, from which the three-gluon Green's function exhibits a more apparent feature. 

A noteworthy final remark is that SDE unquenching techniques, as those developed in refs.~\cite{Aguilar:2012rz,Aguilar:2013hoa}, combined with lattice studies as that of ref.~\cite{Ayala:2012pb}, appear to confirm that the presence of light quarks modifies the ghost and gluon two-point and ghost-gluon three-point Green's functions only slightly at the quantitative level.

\subsection{Generalities}

There have been many past lattice studies~\cite{Boucaud:1998bq,Boucaud:2000nd,Boucaud:2001st,DeSoto:2001qx,Boucaud:2002fx,Boucaud:2003xi} pursuing mainly the computation of the fundamental QCD parameter, $\Lambda_{\rm QCD}$, through a very detailed scrutiny of the running of the three-gluon MOM strong coupling. All these results for the coupling, following \eq{eq:gGammaR}, can be also used to derive the nonperturbative vertex function, as done in ref.~\cite{Athenodorou:2016oyh}. Here, apart from taking advantage of them, we aim towards the completion of the preliminary work of~\cite{Athenodorou:2016oyh}, in which we addressed the calculation of the three-gluon vertex function in the four-dimensional $SU(3)$ gauge theory, by simulating QCD on the lattice with the
tree-level Symanzik improved gauge action (tlSym)~\cite{Weisz:1982zw} which, in addition to the plaquette term
$U^{1\times1}_{x,\mu,\nu}$,  also includes rectangular
$(1\times2)$ Wilson loops $U^{1\times2}_{x,\mu,\nu}$. In particular, the tlSym action reads

\begin{eqnarray}
    S_g =  \frac{\beta}{3}\sum_x\Bigg\{  b_0\sum_{\substack{
      \mu,\nu=1\\1\leq\mu<\nu}}^4\left[1-\re\,\tr\,(U^{1\times1}_{x,\mu,\nu})\right]
     +
    b_1\sum_{\substack{\mu,\nu=1\\\mu\neq\nu}}^4\left[1
    -\re\,\tr(U^{1\times2}_{x,\mu,\nu})\right]\Bigg\},
  \label{eq:Sg}
\end{eqnarray}

\noindent 
where $\beta \equiv 6 / g_0^2$, $g_0$ is the bare lattice
coupling and one sets $b_1=-1/12$ and $b_0=1-8b_1$ as dictated by
the perturbative computation of the improvement coefficient and normalization. The standard Wilson action results from making the choice $b_0=1$ and $b_1=0$ in \eq{eq:Sg}. As will be seen in the next subsection, we have first doubled, in the present study, the number of gauge field configurations for the same two lattice set-ups employed in ref.~\cite{Athenodorou:2016oyh};  produced then one more ensemble with the tlSym action, defining its lattice set-up such as to be left with a physical volume similar to the largest (hence providing reliable estimates for the vertex at equally deep IR momenta) but different lattice spacing; and, finally, obtained three more ensembles of gauge fields simulated with the Wilson action. The idea of including in this detailed analysis an unimproved gauge action originates from 
our intention of guaranteeing that no issue related to the lattice artifacts from the action discretization is relevant for our purposes. All gauge field configurations generated by the above actions are gauge fixed
to the (minimal) Landau gauge.  This is done through the minimization of the following functional [of the $SU(3)$ matrices
$U_\mu(x)$]
\begin{equation}
F_U[g] = \mbox{\re}\left\{ \sum_x \sum_\mu  \hbox{Tr}\left[1-\frac{1}{N}g(x)U_\mu(x)g^\dagger(x+\mu) \right] \right\},
\end{equation}
with respect to the gauge group element $g$.

To get as close as possible to the global minimum, we apply a
combination of an over-relaxation algorithm and Fourier
acceleration, considering the gauge to be fixed when the condition
$|\partial_\mu A_\mu|^2 <10^{-16}$ is fulfilled and the spatial
integral of $A_0$ is constant in time to a relative accuracy better than $10^{-6}$.
Evidently, this procedure cannot avoid the possibility that
lattice Gribov copies are present in the ensemble of gauge fixed
configurations. However, extensive studies in the literature of the quenched case
(see for example~\cite{Bogolubsky:2009dc}) show that such copies
do not seriously affect the qualitative behavior of the Green's functions in question.  

After the lattice configurations have been projected onto the
Landau gauge, the gauge field is obtained as

 \bea 
 A_\mu(x+ \hat \mu/2) = \frac {U_\mu(x) -
 U_\mu^\dagger(x)}{2 i a g_0} - \frac13\, \tr\,\frac{U_\mu(x) -
 U_\mu^\dagger(x)} {2 i a g_0} \;, \label{amu}
 \eea

\noindent 
 with $\hat \mu$ indicating the unit lattice vector in
the $\mu$ direction. It can be then Fourier transformed to momentum space, 

 \bea \widetilde{A}_\mu^a(q)=\frac 1 2 \,\tr\,\sum_x
 A_\mu(x+ \hat \mu/2)\exp[i q\cdot (x+ \hat \mu/2)]\lambda^a \;,
 \label{amufour}
 \eea

\noindent 
where $\lambda^a$ are the Gell-Mann matrices and the trace is evaluated in
color space; and, finally, the two- and three-point gluon Green's
functions are obtained as 

 \bea
 \Delta^{ab}_{\mu\nu}(p) &=& \left\langle \widetilde{A}_{\mu}^{a}(p)\widetilde{A}_{\nu}^{b}(-p)\right\rangle \ ,\nonumber \\
 {\cal G}^{abc}_{\mu\nu\rho}(p,q,r) &=& \left\langle \widetilde{A}_{\alpha}^{a}(p)\widetilde{A}_{\mu}^{b}(q) \widetilde{A}_\nu^c(r) \right\rangle \ ;
\label{greenG}
 \eea

\noindent 
where the $\left\langle \cdot \right\rangle$ stands for a 
Monte-Carlo average replacing here the functional integration over the gauge space. Then, as described in the previous section, one can project out the relevant form factors, $T(p^2)$ and $\Delta(p^2)$, renormalize in the MOM scheme and extract the three-gluon running coupling and the nonperturbative vertex function, following Eqs.~(\ref{eq:g3g}--\ref{eq:gGammaR}). 

To conclude this subsection, we will briefly comment on the role
played by the hypercubic artifacts resulting from the lattice discretization and the consequent breaking of the
$O(4)$ rotational invariance down to the $H(4)$ isometry group. 

The so-called $H(4)$-extrapolation
procedure~\cite{Becirevic:1999uc,Becirevic:1999hj,deSoto:2007ht}
has been proven to cure efficiently the lattice data from these artifacts for the two-point gluon and ghost Green's functions, otherwise plaguing their reliable determination. The procedure basically works as follows, any dimensionless correlation function (as $q^2\Delta$ and $q^6 T$) evaluated on the lattice must depend on the (dimensionless) lattice momentum
$a\,{q}_\mu$, where
 \bea q_\mu = \frac{2\pi n_\mu}{N_\mu a} \; , \qquad
 n_\mu=0,1,\dots,N_\mu \; ,
 \eea $N_\mu$ being the number of lattice
sites in the $\mu$ direction (in our case, $N_\mu=N$ for all $\mu$). Since $O(4)$ is broken down to
$H(4)$, it depends not only on $a^2q^2$ but also on $a^2q^{[4]}/q^2$, at the first order, where $q^{[4]}=\sum_\mu q_\mu^4$ is the first $H(4)$-invariant. All the different configurations of lattice momenta, $q_\mu$, taking the same value for all the $H(4)$-invariants are obviously invariant under $H(4)$ transformations, and constitute a so-called $H(4)$-orbit. However, different $H(4)$-orbits may take the same value for $q^2$, differing only by  $q^{[4]}$ or higher-order invariants. Thus, a correlation function evaluated on the lattice at momenta belonging to those different $H(4)$-orbits will differ from each other but must take the same continuum value. This is an apparent manifestation of the $O(4)$-breaking, in order to be dealt with, the standard recipe was to introduce a kinematical cut in lattice momenta configuration. The purpose was to exclude those carrying a momentum component much larger than the others, intended to only retaining small $a^2q^{[4]}/q^2$ contributions; and subsequently perform an average over all the $O(4)$-invariant momenta configurations. Instead, in the $H(4)$-extrapolation, one only averages over any combination of momenta within the same $H(4)$ orbit and extrapolates then the results towards the continuum limit where the effect of $a^2 q^{[4]}$ vanishes. 

An extension of the $H(4)$-extrapolation procedure has been also developed and applied to deal with the hypercubic artifacts of the three-gluon Green's functions in ref.~\cite{Boucaud:2013jwa}. Therein, the results obtained from $H(4)$-extrapolation and from the standard  $O(4)$-average were compared and their differences appeared only visible for lattice momenta such that $\sqrt{a^2q^2} \ge \pi/4$. Here, we have also applied $H(4)$-extrapolation and $O(4)$-average and only retained those lattice momenta which lead to results that compare well after each of the two procedures has been applied.

\subsection{Results}

Before displaying the lattice results that we obtained and analyzed in this work, we give first the different lattice set-ups for all the ensembles produced and exploited here. The simulations parameters are reported in Tab.~\ref{Tab:set-up}. As can be seen there, 
we have implemented the two different discretized gauge actions, Wilson and tlSym, described in the previous subsection, adopting three set-ups for each, corresponding to five different lattice bare couplings (lattice spacings), $g_0^2=6/\beta$, their physical volumes ranging from $4.45^4$ to $15.6^4$ fm$^4$. For the sake of comparison, we have also exploited the lattice data for the three-gluon running coupling, previously published and analyzed in investigations mostly addressed to the calculation of $\Lambda_{\rm QCD}$~\cite{Boucaud:1998bq,Boucaud:2000nd,Boucaud:2001st,DeSoto:2001qx,Boucaud:2002fx,Boucaud:2003xi}, and derived from them the nonperturbative vertex function. These last data were then obtained from many different lattice simulations, generated with the Wilson gauge action at several $\beta$'s ranging from 5.6 to 6.0, and physical volumes from $2.4^4$ to $5.9^4$ fm$^4$.

\begin{table}[htb]
\begin{tabular}{|c||c|c|c||c|c|c|}
\hline
$\beta$ & 4.2 & 3.8 & 3.9 & 5.8 & 5.6 & 5.6 \\
\hline 
$N$ &  32 & 48 & 64 & 48 & 48 & 52 \\
\hline 
(Volume)$^{1/4}$ [fm] & 4.45 & 13.7 & 15.6 & 6.72 & 11.3 & 12.3 \\
\hline
confs. & 420 & 1050 & 2000 & 960 & 1920 & 1790 \\
\hline 
action & tlSym & tlSym & tlSym & Wilson & Wilson & Wilson \\
\hline
\end{tabular}
\caption{\small Lattice set-up parameters for the simulations employed here using either the Wilson or tree-level Symanzik (tlSym) discretized gauge actions. The lattice bare coupling, for the case of $SU(3)$, is given by $g_0^2=6/\beta$, $N$ stands for the number of lattice sites in any of the four dimensions, the physical volume is obtained from the lattice spacings taken from  refs.~\cite{Guagnelli:2000jw}, for the Wilson action, and~\cite{Athenodorou:2016gsa} for tlSym. The number of the exploited configurations of gauge fields is also given in the fourth row.}
\label{Tab:set-up}
\end{table}

\begin{figure}[t!]
\begin{tabular}{cc}
\includegraphics[width=8cm]{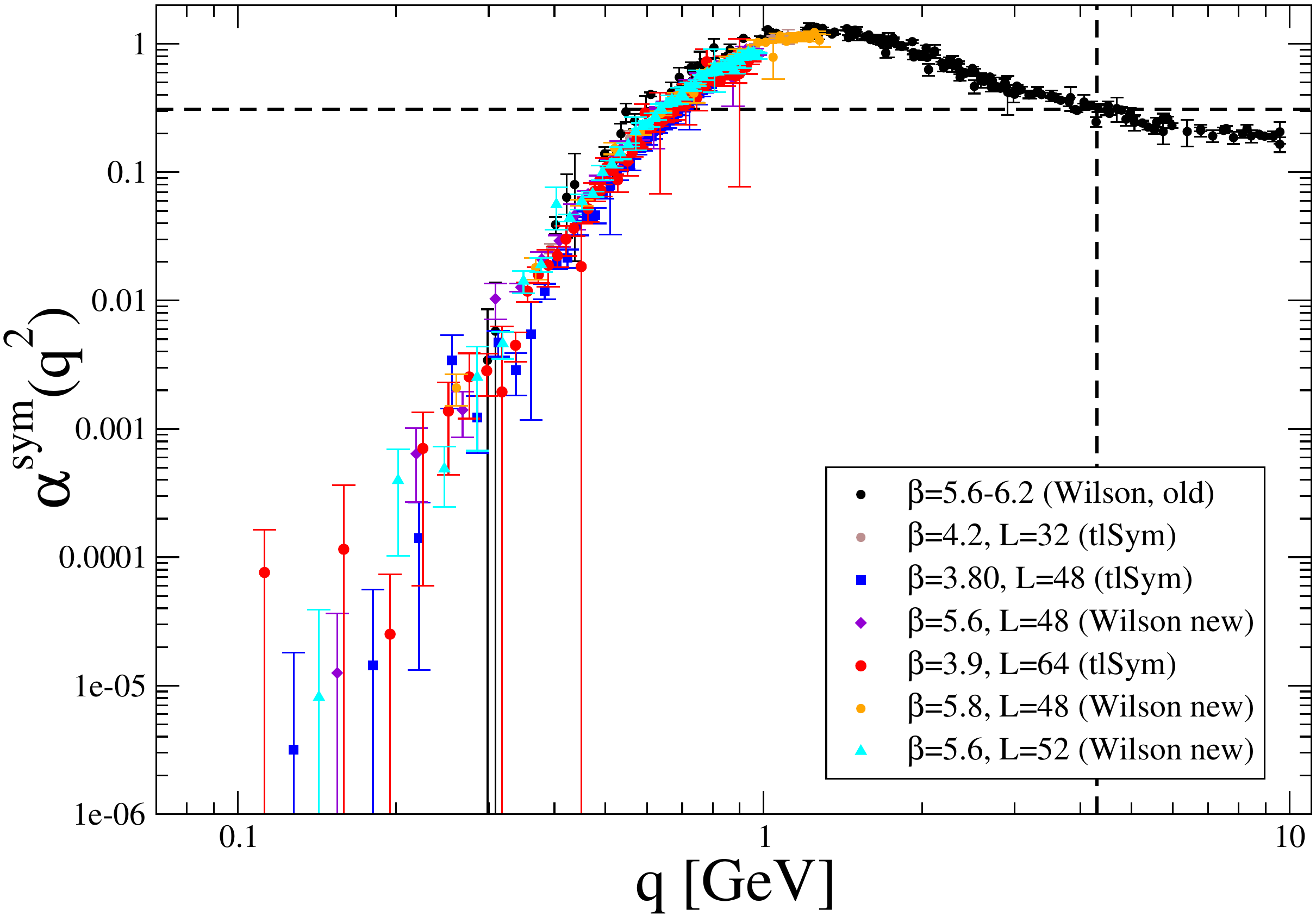} 
&\includegraphics[width=8cm]{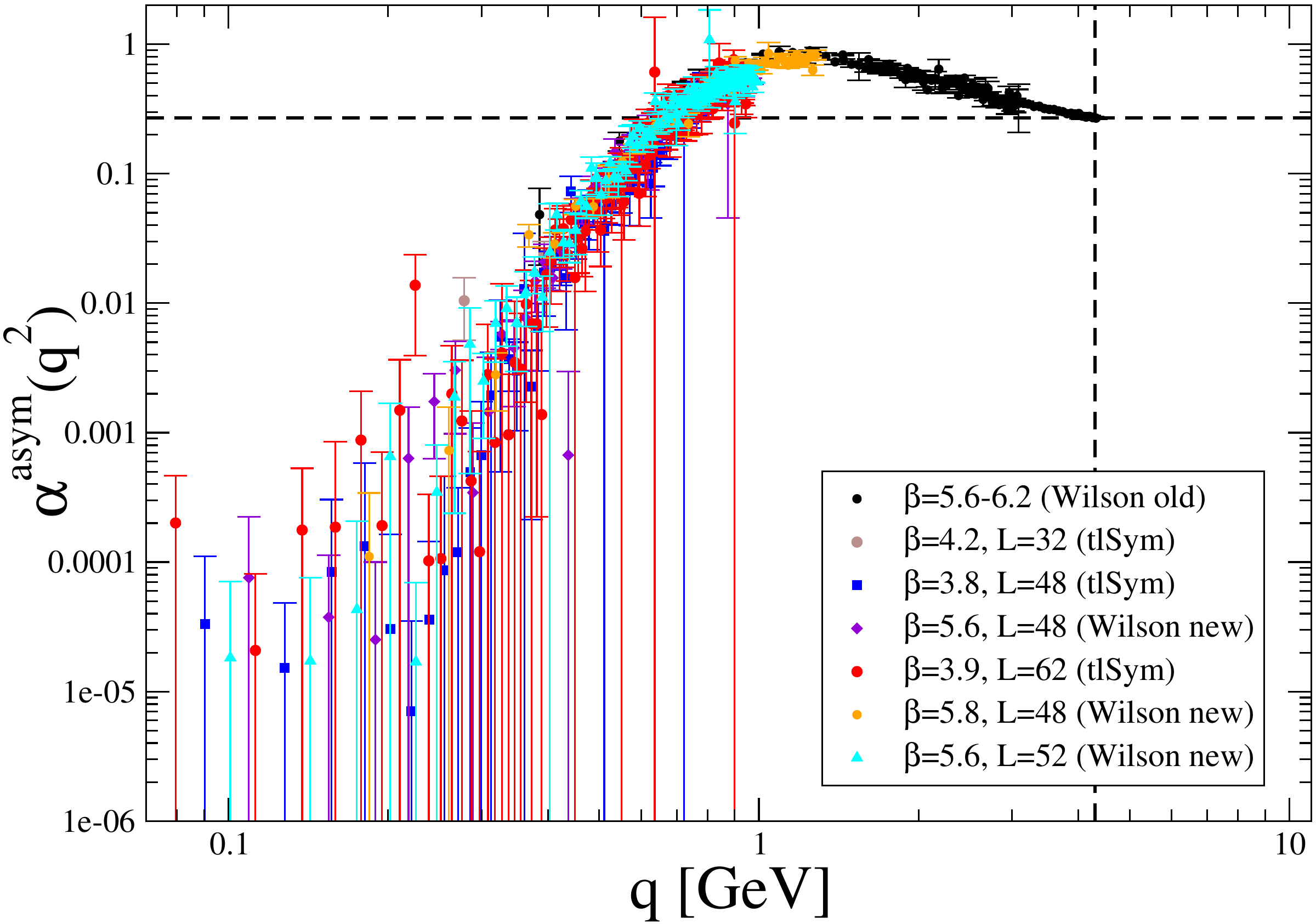} 
\end{tabular}
\caption{\label{fig:g3g} \small The strong running coupling, \eq{eq:alpha}, in both symmetric (left panel) and asymmetric (right panel) MOM schemes, where $g_R$ is computed by using, respectively, Eqs.~\eqref{eq:g3g} and \eqref{eq:g3gasym}. We use logarithmic scales for both axes in both panels. Data obtained from the six set-ups reported in Tab.~\ref{Tab:set-up} appear displayed with different symbols and colors (as shown in the legends) and, additionally, data previously published and investigated in refs.~\cite{Boucaud:1998bq,Boucaud:2000nd,Boucaud:2001st,DeSoto:2001qx,Boucaud:2002fx,Boucaud:2003xi} are plotted (black solid circles), for the sake of comparison and to cover the UV region where we estimate $\alpha^{\rm sym}(4.3~\mbox{\rm GeV})=0.31$ and $\alpha^{\rm asym}(4.3~\mbox{\rm GeV})=0.27$ (dashed lines). }
\end{figure}

Within the approach we follow, the road to the computation of the 1-PI renormalized vertex function implies, in a first step, to obtain the three-gluon running coupling defined in Eqs.~(\ref{eq:g3g}) and (\ref{eq:g3gasym}), respectively, for symmetric and asymmetric MOM schemes. This calculation can be directly performed by the implementation of the bare form factors, $\Delta(p^2)$ and $T(p^2)$, projected out from the lattice two- and three-point Green's functions through Eqs.~(\ref{eq:prop}),(\ref{eq:projsym}) and (\ref{eq:projasym}). A main gain of proceeding so is that all the dependence on the cut-off regularization parameters should cancel for the ratios of bare quantities in the definitions of strong coupling; and so can be explicitly examined. As far as one properly deals with the lattice discretization artefacts, the outcome of Eqs.~(\ref{eq:g3g},\ref{eq:g3gasym}) would only be a function of the squared momentum at the subtraction point in either the symmetric or the asymmetric configurations, irrespectively of what set of lattice parameters is used. A striking verification of the latter can be seen in the two plots of Fig.~\ref{fig:g3g}, where the lattice results for 

\begin{equation}\label{eq:alpha}
\alpha^{i}(p^2) \ = \ \frac{\left[g_R^{i}(p^2)\right]^2}{4 \pi} \ ,
\end{equation}

\noindent
with $g_R^{i}$ given by Eqs.~(\ref{eq:g3g}) and (\ref{eq:g3gasym}), respectively, for $i=$sym and $i=$asym. The strong coupling data obtained from all the lattice ensembles, with very different bare couplings, lattice spacings and physical volumes, appear to be very well on top of each other when their window of momenta overlap (where the lattice artifacts happen to be under control). It is worth emphasizing that no renormalization or rescaling factor is needed: the physical scaling shown by the plots of Fig.~\ref{fig:g3g} directly results from the evaluation of Eqs.~(\ref{eq:g3g},\ref{eq:g3gasym}) with bare lattice inputs~\footnote{No direct rescaling is needed for the coupling, however we have corrected by a 5 \% --admitting that relative error-- the value of the lattice spacing in physical units for the Wilson-action set-up at $\beta=5.8$, with the criterion of obtaining an optimal scaling.}.

\begin{figure}[t!]
\begin{tabular}{cc}
\includegraphics[width=8cm]{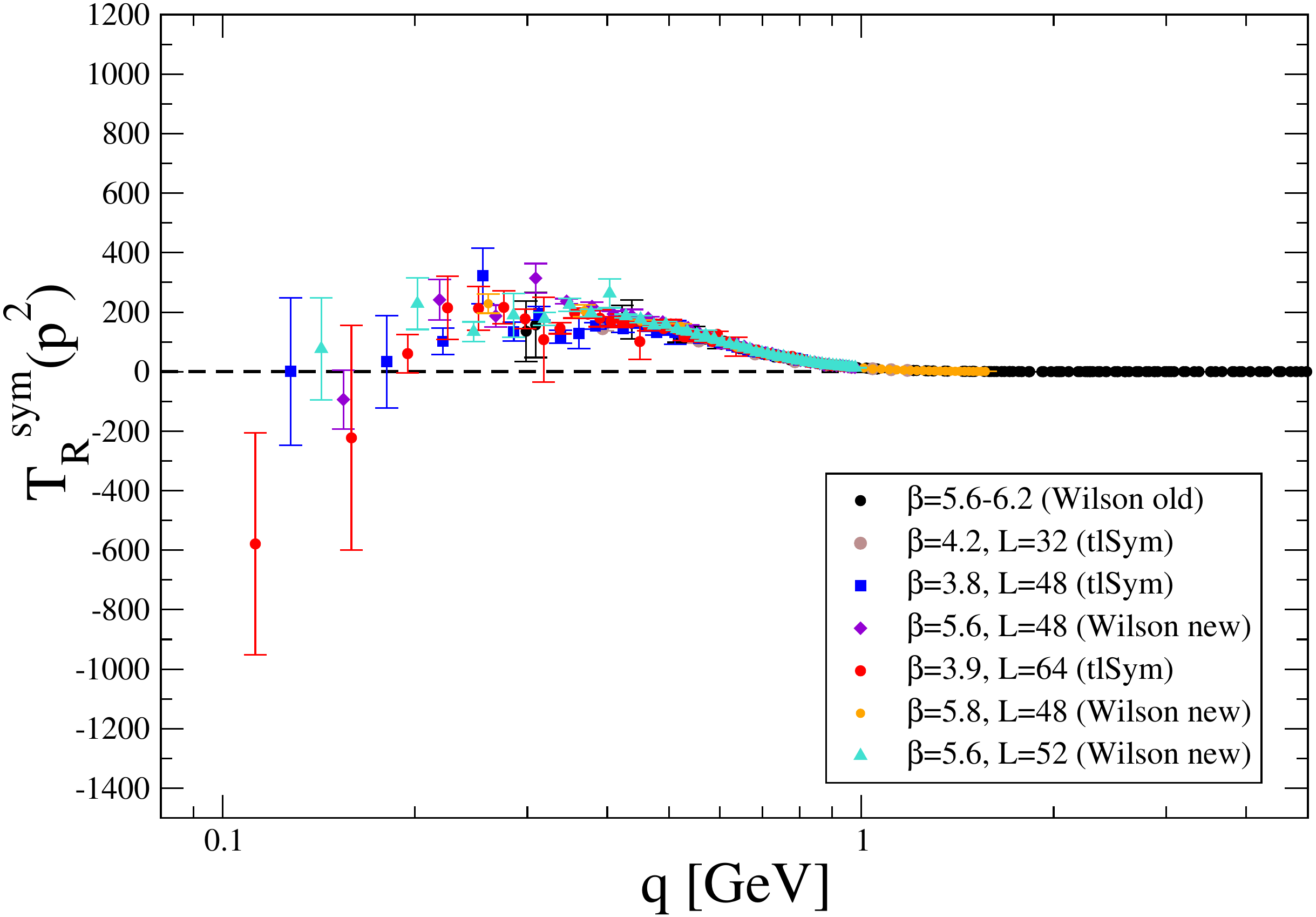} 
&\includegraphics[width=8cm]{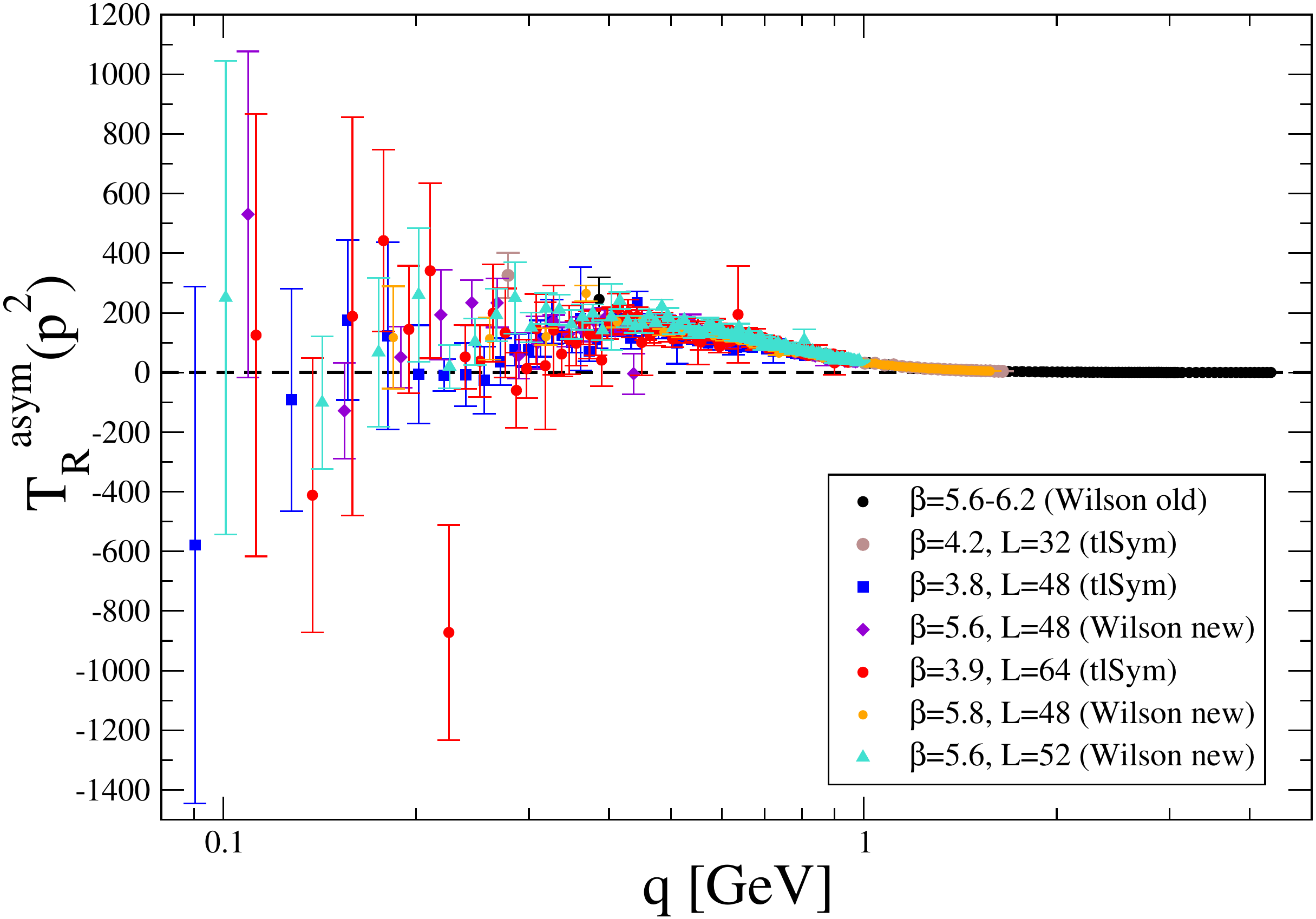} 
\end{tabular}
\caption{\label{fig:TRs} \small The three-gluon form factor, renormalized at $\mu=4.3$ GeV according to \eq{eq:renT}, after being projected out through Eqs.~(\ref{eq:projsym}) and (\ref{eq:projasym}) from, respectively, the symmetric (left panel) and asymmetric (right panel) lattice bare Green's functions.}
\end{figure}

Moreover, we have also computed the renormalized three-gluon form factor, $T^{i}_R$, following \eq{eq:renT}, and displayed the results in Fig.~\ref{fig:TRs}. For this and, in the following, for all the renormalized quantities, we have chosen  $\mu=4.3$ GeV as the subtraction momentum. Precisely, the use of previously published lattice data together with the new ones produced here made possible to enlarge the window of momenta for reliable estimates, covering a wider region, containing large UV momenta, where this renormalization point is included. In particular, as it appears indicated by the dashed lines in both panels of Fig.~\ref{fig:g3g}, we thus get: $\alpha^{\rm asym}(4.3~\mbox{\rm GeV})=0.31$, for the symmetric case, and $\alpha^{\rm asym}(4.3~\mbox{\rm GeV})=0.27$ for the asymmetric one. The results for the renormalized form factor in the symmetric case robustly show a change of sign and, hence, a zero crossing lying somewhere in between $0.1$ and $0.2$ GeV. In the asymmetric case, the results are statistically noisier but clearly consistent with the same feature. 

\begin{figure}[thb!]
\begin{tabular}{cc}
\includegraphics[width=8cm]{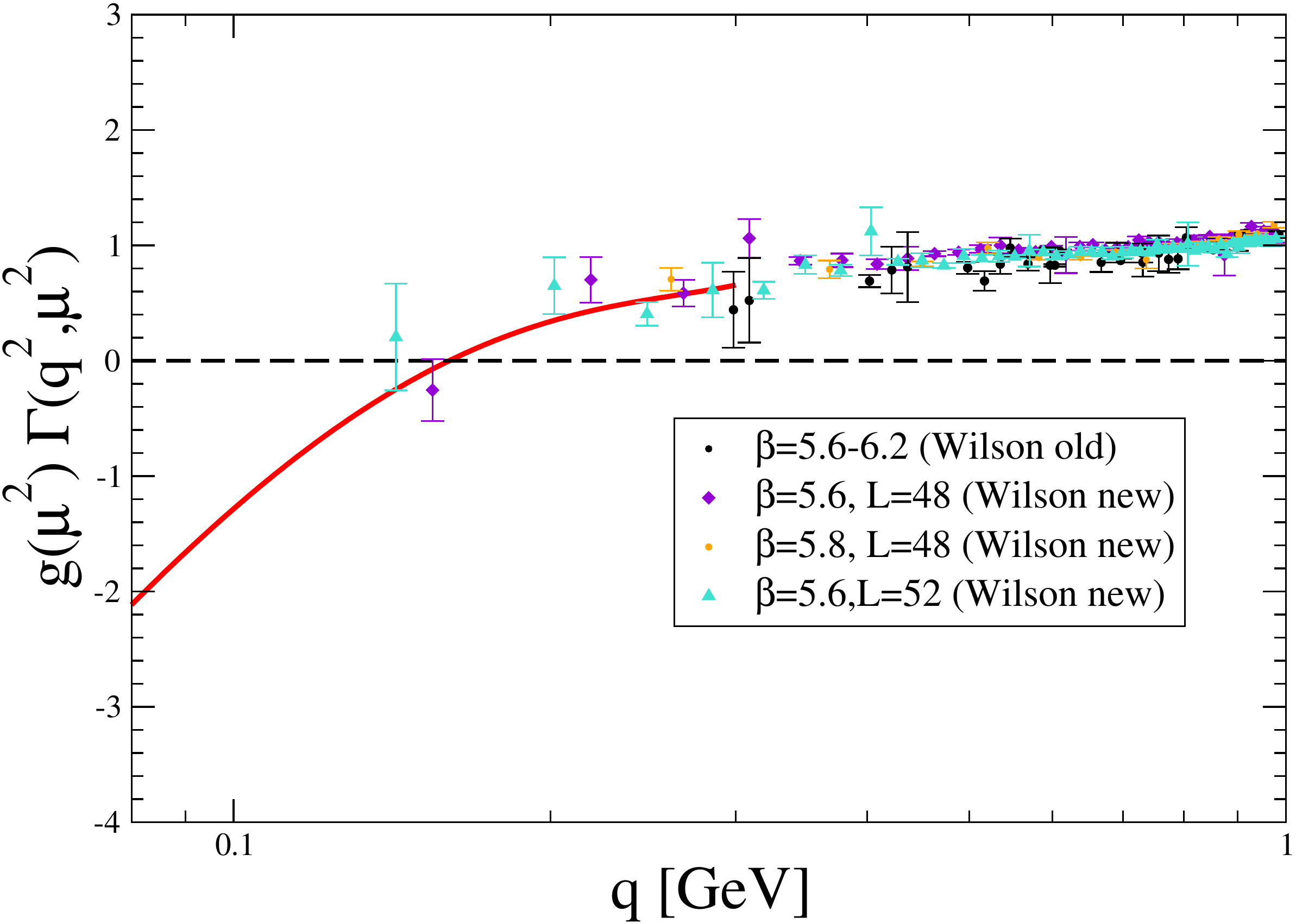} & 
\includegraphics[width=8cm]{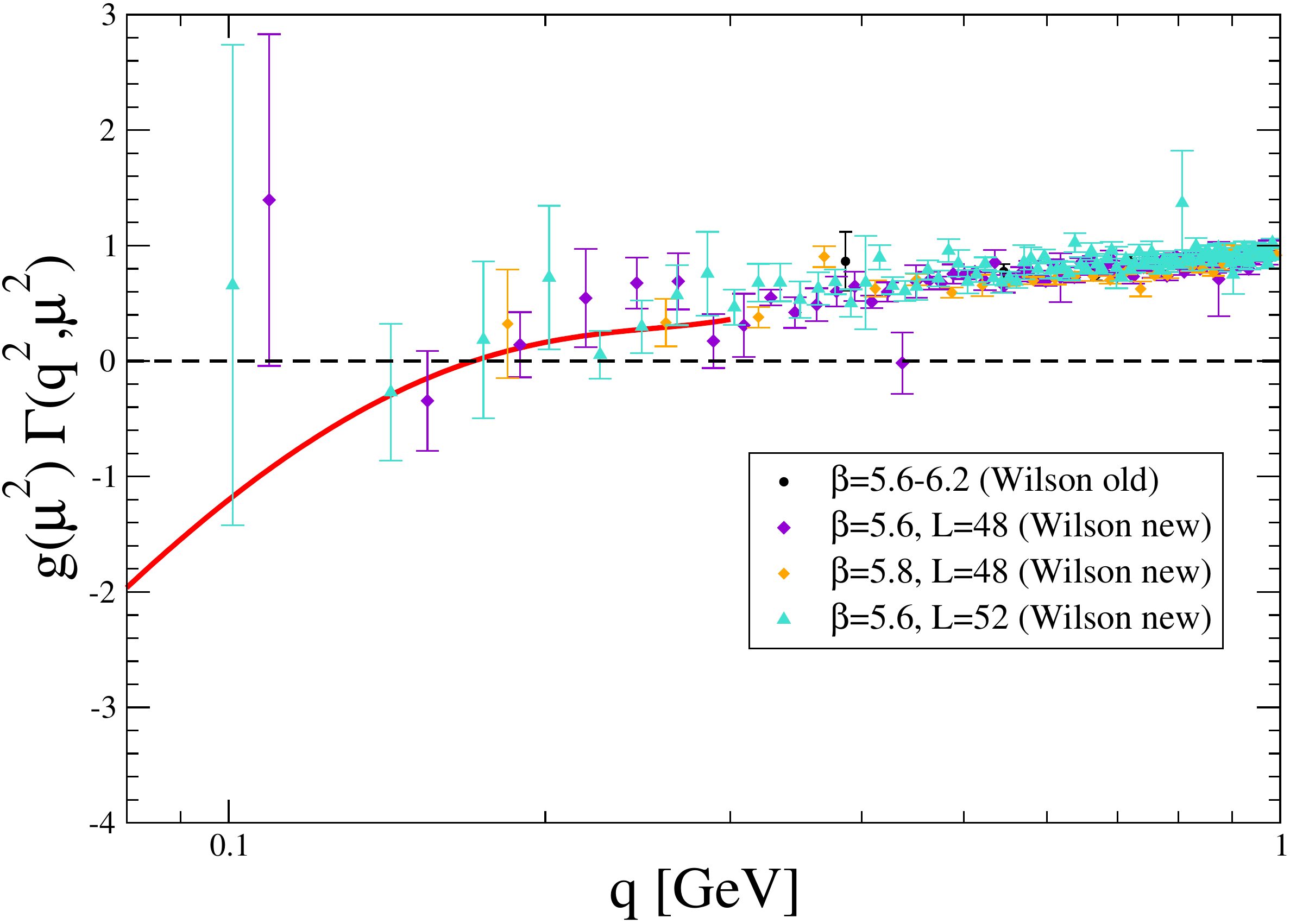} \\
\includegraphics[width=8cm]{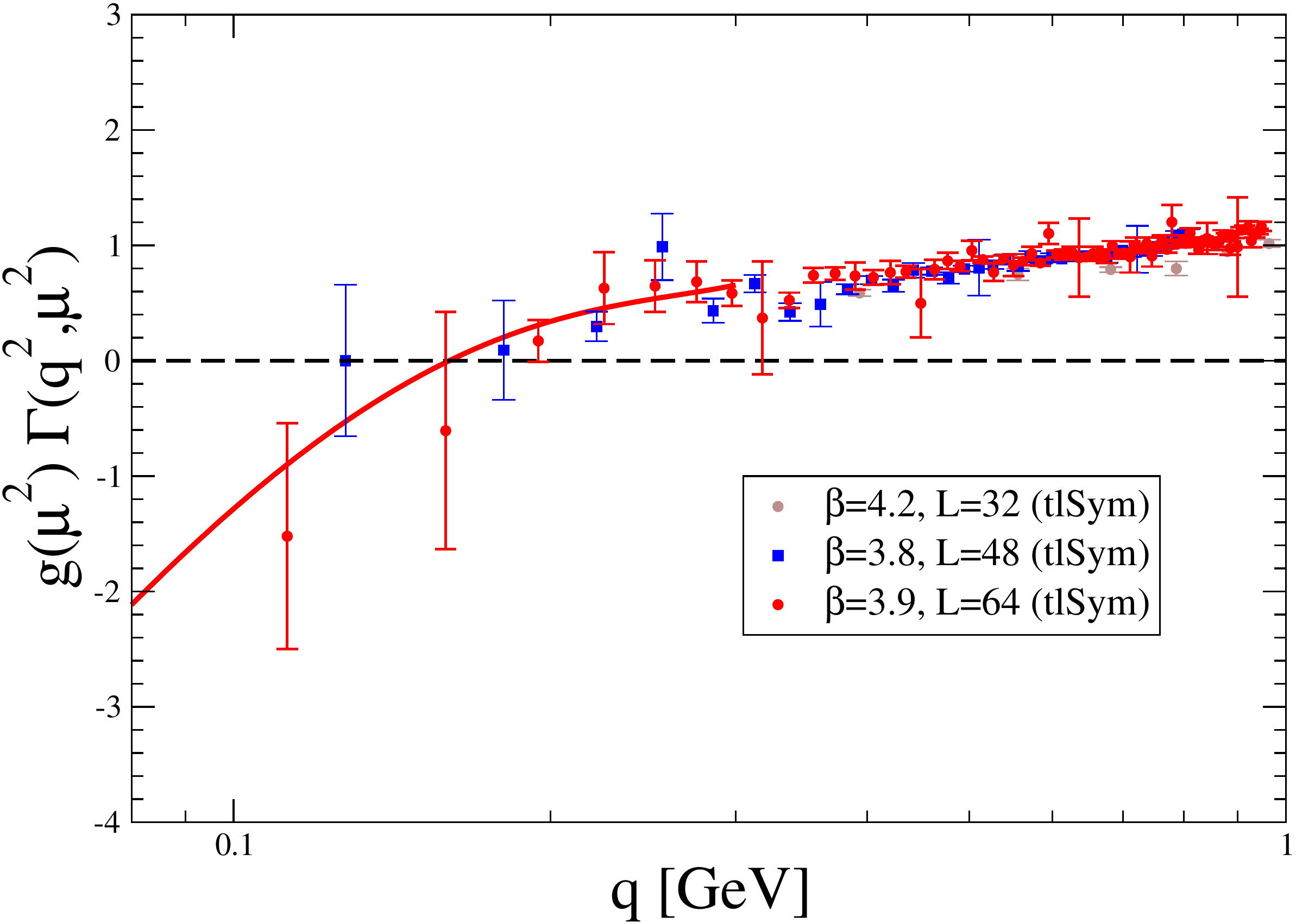} & 
\includegraphics[width=8cm]{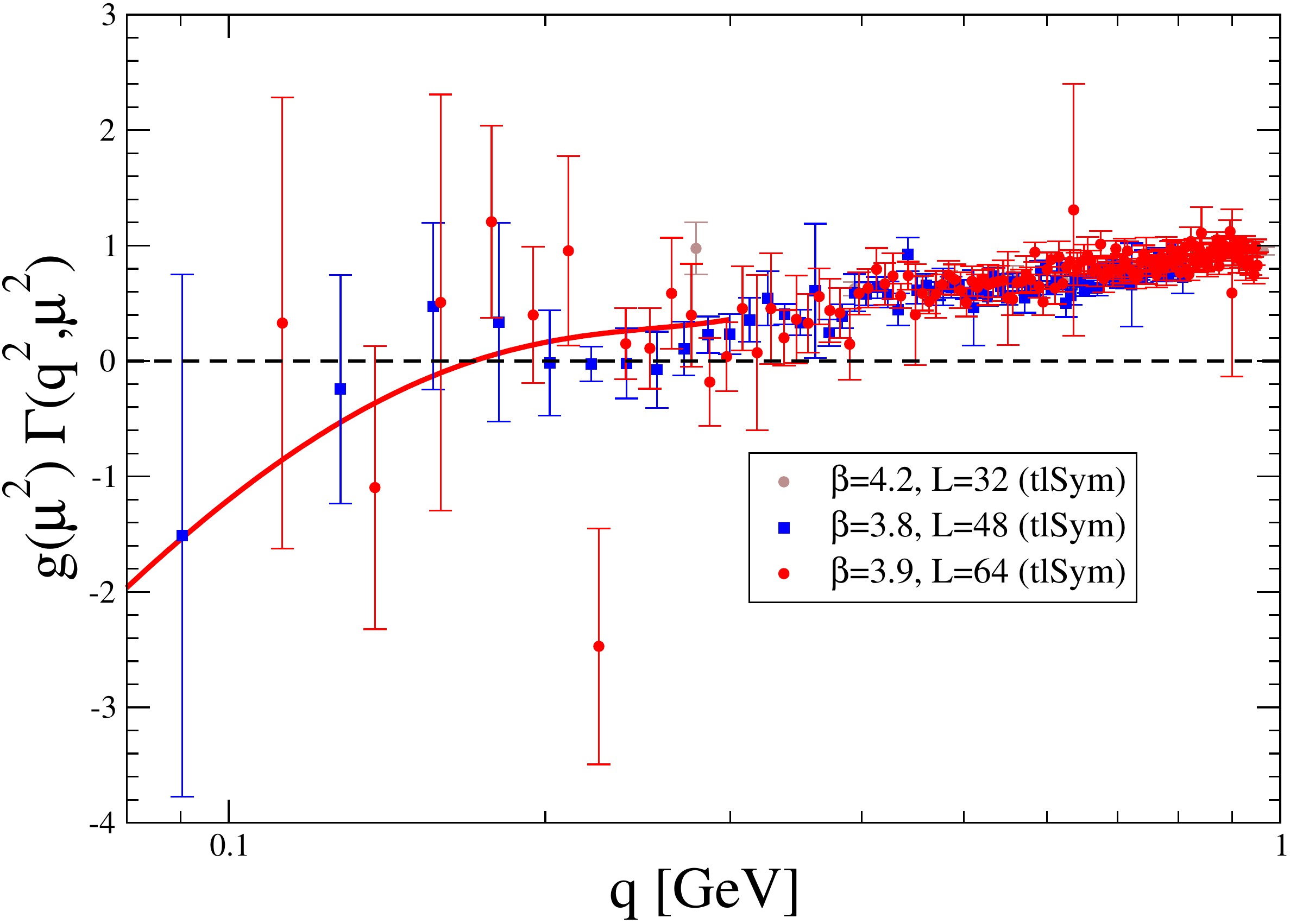}
\end{tabular}
\caption{\label{fig:1PI} \small  Three-gluon 1-PI form factors obtained from lattice results simulated with the Wilson (upper panels) and the tlSym (lower panels) gauge actions, plotted in terms of the momenta displayed in logarithmic scale, for symmetric (left) and asymmetric (right) momenta configuration.  The renormalization point is $\mu$=4.3 GeV. The red solid lines result from SDE-based fits which will be explained in the next section and that are included here for comparative purposes.}
\end{figure}

Finally, in order to obtain the 1-PI vertex functions, we only need to resort to their connection to the strong coupling which results, in both the symmetric and the asymmetric momenta configurations, from \eq{eq:gGammaR}. As this equation reads, the calculation involves a gluon two-point Green's function renormalized at the chosen momentum which, precisely, carries all the renormalization-point dependence for the vertex function. This two-point Green's function being a renormalized quantity, on the understanding of which exists now a clear consensus and which appears to be very accurately known, we can take for it the results obtained with the best available gluon propagator lattice data in the literature; namely, those simulated with the largest physical volume~\cite{Bogolubsky:2009dc}. We thus only need to combine, according to \eq{eq:gGammaR}, the renormalized results for the two-point function with those for the three-gluon strong coupling, shown in Fig.~\ref{fig:g3g}, directly obtained from bare lattice Green's functions. Or one can apply instead Eqs.~\eqref{eq:GammaR} and \eqref{eq:GammaRasym} and use there the results for the renormalized three-gluon form factors collected in Fig.~\ref{fig:TRs}. In both cases, we will be left with the results for the 1-PI vertex functions, which appear displayed in Fig.~\ref{fig:1PI}. 

In Fig.~\ref{fig:1PI}, we have grouped only the results obtained with the same gauge action, either Wilson (upper panels) or tlSym (lower panels). In both cases, the zero-crossing feature clearly happens for the symmetric momenta configuration (left) and is remarkably consistent with the (noisier) asymmetric configuration (right). Indeed, the Wilson-action symmetric-configuration data also appear to offer a robust confirmation for the happening of the zero crossing~\footnote{This is in agreement with the very recent claim for further evidences for the zero crossing in the asymmetric three-gluon vertex~\cite{Duarte:2016ieu}}. In the aim of guiding the eye for comparative purposes, we have also included, in all the plots, the SDE-inspired fits that will be further discussed in the next section. It is not worthless emphasizing that, even if one is willing to analyze independently any of the four lattice set-ups producing estimates for momenta lying on the zero crossing region ($\beta$=5.6 for 48$^4$ and 52$^4$ lattices, in the Wilson case, and $\beta$=3.8 and $\beta$=3.9, tlSym), the attained conclusions would be plainly compatible with a global analysis, as the data from any of these set-ups appear to be consistent with each other and behave as expected, if the SDE interpretation for the zero crossing is correct.

\section{SDE-based analysis}
\label{sec:SDE}

As was preliminary shown in~\cite{Athenodorou:2016oyh}, lattice data for the three-gluon 1-PI vertex functions, in particular the change of sign and the zero crossing they exhibit in the deep IR, appear to be consistent with a noteworthy effect that had been previously shown in literature (notably in \cite{Aguilar:2013vaa,Tissier:2011ey}); namely, the (4-d) logarithmic singularity induced in the three-gluon vertex function by the {\it nonperturbative} ghost loop diagram  contributing to the gluon propagator SDE. Specifically, employing a nonperturbative Ansatz for the ghost-gluon vertex that satisfies the appropriate Slavnov-Taylor identity (STI), one is left with the following IR contribution to the gluon self-energy for this ghost-loop diagram~\cite{Aguilar:2013vaa}, 
\begin{equation}
\label{eq:PIc}
\Pi_c(p^2) \ = \ \frac{g^2 C_A}{6} p^2 F(p^2) \int_k \frac{k^2}{k^2 (k+p)^2} \ , 
\end{equation} 
where $C_A$ is the Casimir eigenvalue in the adjoint representation, $F(p^2)$ is the ghost dressing function and $\int_k \equiv \mu^\varepsilon/(2\pi)^d \int d^dk$ is the dimensional regularization measure, with $d=4-\varepsilon$ and $\mu$ is the 't Hooft mass. The above leading contribution in the vanishing momentum limit, $p^2 \to 0$, evidently behaves like $p^2 \ln{p^2/\mu^2}$ and leads to the following very accurate parametrization, 
\begin{equation}\label{eq:Deltam1}
\Delta^{-1}_R(p^2;\mu^2) \underset{p^2/\mu^2 \ll 1}{=}p^2\left[a+b\ln\frac{p^2+m^2}{\mu^2}+c\ln \frac{p^2}{\mu^2} 
	\right]+m^2 \ ,
\end{equation}  
for the IR form of the gluon propagator which emerges from its complete SDE; with $a,b,c$, and $m^2$ suitable renormalization-dependent parameters capturing explicitly both the effect of the ghost loop sketched by \eq{eq:PIc} and the finiteness of the gluon propagator at vanishing momentum, $\Delta_R^{-1}(0;\mu^2)=m^2$. Note that this finiteness is the consequence of an effective mass being acquired which {\it protects} the gluon loops against the logarithmic singularities resulting from the ghost loops~\footnote{Then, although the ghost is directly transparent to the mass generation mechanism, the absence of additional singularities resulting from the nonperturbative gluon loops, which owes to the gluon mass generation, guarantees the finiteness of the ghost dressing function at vanishing momentum.}. 

\begin{figure}[thb!]
\begin{tabular}{cc}
\includegraphics[width=8cm]{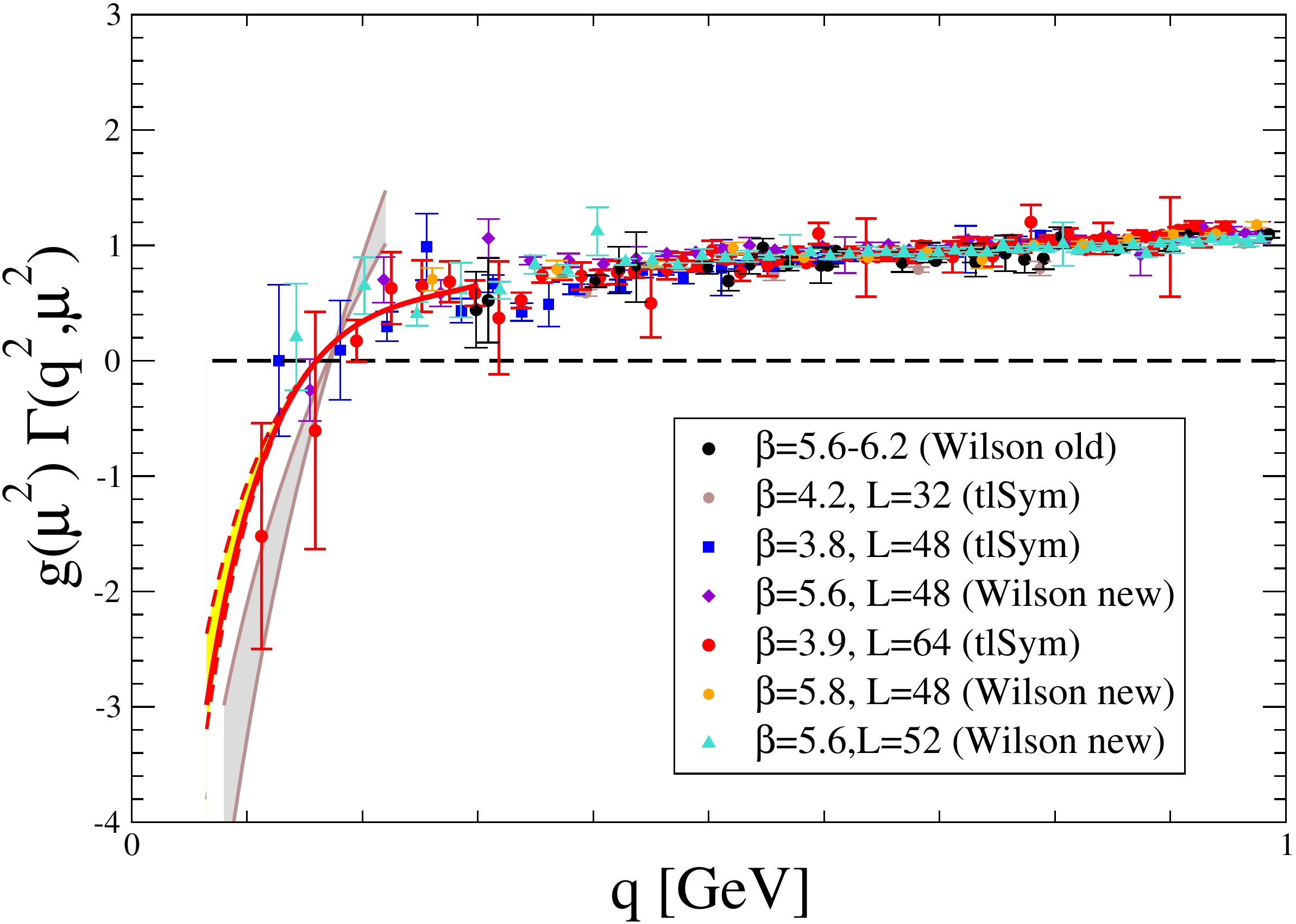} & 
\includegraphics[width=8cm]{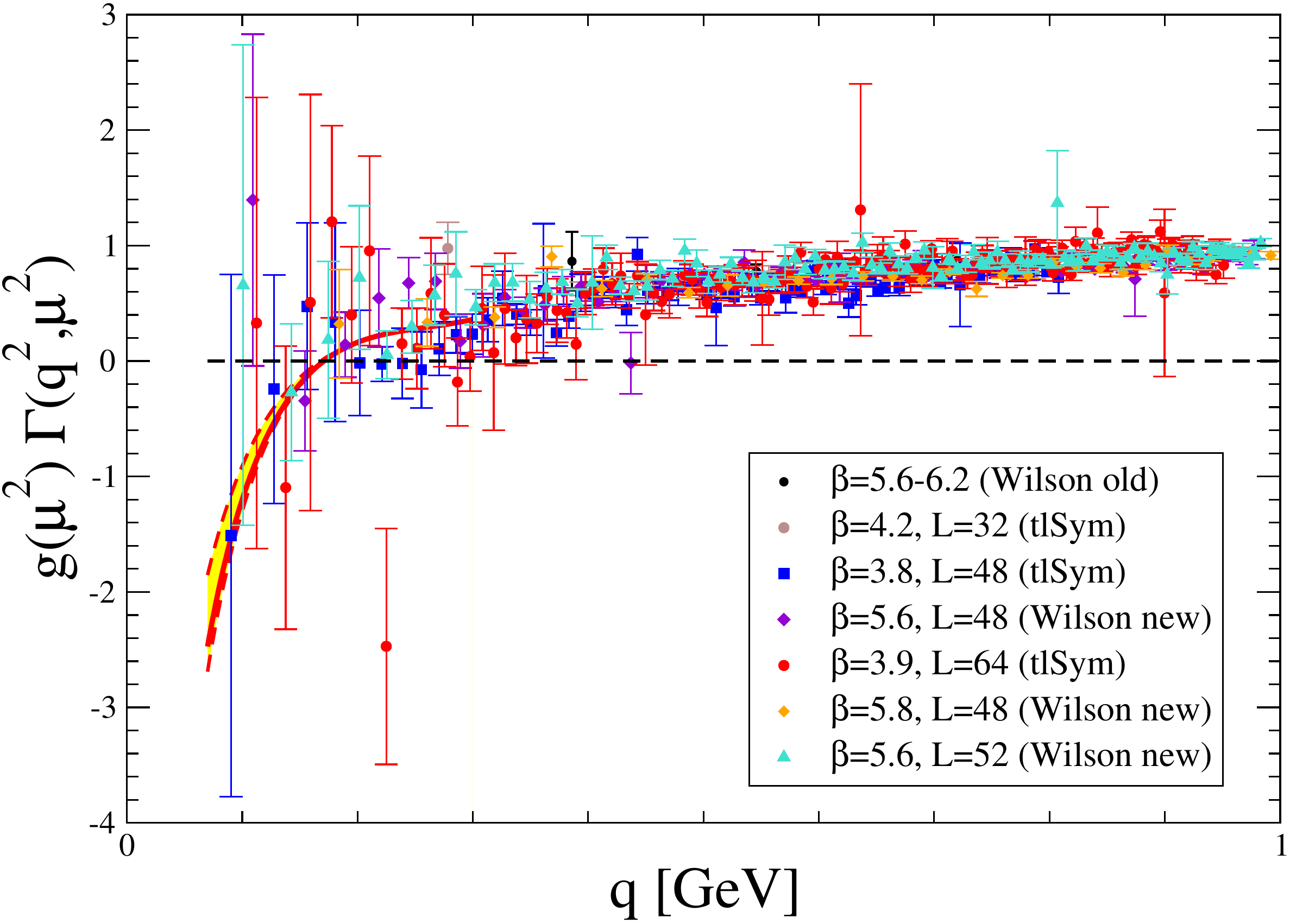} 
\end{tabular}
\caption{\label{fig:gGamma} \small  The same three-gluon 1-PI form factors displayed in ref.~\ref{fig:1PI}, plotted now all together and making use of a linear scale, for symmetric (left) and asymmetric (right) momenta configuration.  The renormalization point is $\mu$=4.3 GeV. The red solid and dashed lines result from the best fits with \eq{eq:gGammaF}, while the brown solid lines correspond to \eq{eq:gGamma0}, as explained in the text. Yellow and brown bands depict the uncertainty resulting from the range of $c$ estimated in ref.~\cite{Athenodorou:2016gsa}, in the cases of, respectively, \eq{eq:gGammaF}'s and \eq{eq:gGamma0}'s fits.}  
\end{figure}

Any standard Green's function can be related to the same one with {\it background} legs, within the PT-BFM approach, by the use of the so-called ``{\it background quantum}" identities~\cite{Grassi:1999tp,Binosi:2002ft,Binosi:2009qm}). The ones with {\it background legs}, when projected according to Eqs.~(\ref{eq:projsym},\ref{eq:projasym}) and by virtue of the Abelian STI that the PT-BFM propagators are constructed to obey, will be led in the IR by the derivative of the inverse of the gluon propagator, represented by \eq{eq:Deltam1}~\cite{Aguilar:2013vaa}. Thus, the three-gluon 1-PI form factors derived from the background Green's functions can be proven to behave in the deep IR as
\begin{eqnarray}
\Gamma^{i,(B)}_{T,R}(p^2;\mu^2) &\underset{p^2/\mu^2\ll 1}{\simeq}&  F_R(0;\mu^2) \frac{\partial}{\partial p^2}  \Delta^{-1}_R(p^2;\mu^2) \ + \ \dots \nonumber \\ 
& \simeq & F_R(0;\mu^2) \left( a + b\ln\frac{m^2}{\mu^2} + c \right) + c \  F_R(0;\mu^2) \ln \frac{p^2}{\mu^2}  \ + \ \dots \ ;  
\label{eq:GammaB}
\end{eqnarray}
where $F_R(0;\mu^2)$ is the renormalized ghost dressing function evaluated at zero momentum and where the dots stand for subleading corrections that, as discussed in \cite{Athenodorou:2016gsa}, might be collectively taken into account by adding an extra constant term which, contrarily to the leading contribution, depends a priori on the momenta configuration. On the other hand, the connection between the background and the standard vertex functions is controlled by the ghost-gluon dynamics and will essentially introduce a finite correction not modifying the leading logarithmic divergence in \eq{eq:GammaB}. Thus, we can eventually write
\begin{equation}\label{eq:gGamma0}
g_R^{i}(\mu^2) \Gamma^{i}_R(p^2;\mu^2) \ = \ a_{ln}^{i}(\mu^2) \ln \frac{p^2}{\mu^2} \ + \ a_0^{i}(\mu^2)  \ + \ o(1) \ ,
\end{equation}
where $a_0^i(\mu^2)$ is a constant which borrows from all the subleading corrections and that will be considered, in the following, as a free parameter to be fitted; while the logarithmic slope, $a_{ln}^i(\mu^2)=g_R^i(\mu^2) \ c \ F_R(0;\mu^2)$, is known from gluon and ghost two-point Green's functions and from the value of the three-gluon coupling at the renormalization point. Indeed, at $\mu=$4.3 GeV, we have $F_R(0,\mu)\simeq 2.9$ from ref.~\cite{Bogolubsky:2009dc}, $c \in (0.35,0.55)$ from ref.~\cite{Athenodorou:2016gsa} (where the range is thought to account for the uncertainty coming from the two fits of \eq{eq:Deltam1} to data from $80^4$ and $96^4$ lattices at $\beta$=5.7~\cite{Bogolubsky:2009dc}) and have estimated the coupling from the data displayed in Fig.~\ref{fig:g3g} for both symmetric and asymmetric configurations. We thus obtain the values for $a_{ln}^i$ reported in Tab.~\ref{Tab:param} (first or second column, for the symmetric case, and fourth or fifth, for the asymmetric) and used in Fig.~\ref{fig:gGamma}.

However, \eq{eq:gGamma0}, with values for $a_{ln}^i$ lying inside its predicted range, can hardly account for data within a fitting window of momenta up to 0.2 GeV, in particular for the symmetric case (see the brown band in the left plot of Fig.~\ref{fig:gGamma}). Otherwise said, a fit of~\eq{eq:gGamma0} with both $a_0^i$ and $a_{ln}^i$ as free parameters would yield an optimal best-fit result for the latter about twice smaller than the lowest value reported in Tab.~\ref{Tab:param}. This seems to suggest that the first subleading correction, at least, is needed to describe properly the data around 0.2 GeV. As we know from refs.~\cite{Boucaud:2010gr,RodriguezQuintero:2011vw,Binosi:2016xxu}, the first subleading correction introduced by the ghost-gluon dynamics should behave as $p^2 \ln(p^2)$. We thus correct \eq{eq:gGamma0} as
\begin{equation}\label{eq:gGammaF}
g_R^i(\mu^2) \Gamma^i_R(p^2;\mu^2) \ = \ a_{ln}^i(\mu^2) \ln \frac{p^2}{\mu^2} \ + \ a_0^i(\mu^2) \ + \ 
a_2^i(\mu^2) \ p^2 \ln\frac{p^2}{M^2} \ + \ o(p^2) \ ,
\end{equation}
where, again, $a_2^i$ and $M$ will be free parameters capturing subleading contributions. $M$ differs a priori from the renormalization point since it is absorbing the ${\cal O}(p^2)$-contribution which, for the sake of consistence, is also required. Then, by applying \eq{eq:gGammaF} with $a_{ln}^i$ also as a free parameter, we get the best fits displayed in solid red line in the left panels (for the symmetric case) of Figs.~\ref{fig:1PI} and \ref{fig:gGamma}. These fits give a nice description of all data below 0.3 GeV and correspond to the best-fit parameters reported in Tab.~\ref{Tab:param} (third column). In particular, the optimal value for $a_{ln}^{sym}(\mu^2)$, combined with $F_R(0,\mu^2)$ and $g^{sym}_R(\mu^2)$, allows for a prediction of the coefficient $c$ controlling the massless ghost loop contribution, 
\begin{equation}\label{eq:agree}
c \ \simeq \ 0.48 \ , 
\end{equation}
which happens to agree pretty well with the range of results obtained in ref.~\cite{Athenodorou:2016oyh} from gluon propagator lattice data given in~\cite{Bogolubsky:2009dc}, namely $ c \in \ (0.35,0.55)$. Next, we can take this best-fit estimate of $c$, obtained from the symmetric configuration, to determine $a_{ln}^{asym}$, apply it in \eq{eq:gGammaF} and produce then the best fits for the asymmetric configuration displayed in the right panels of Fig.~\ref{fig:1PI} and \ref{fig:gGamma}. In the view of all the plots and, specially, of \eq{eq:agree}, it can be concluded that the lattice three-gluon vertex data we produced and introduced in the previous section (part of which were published and analyzed in \cite{Athenodorou:2016oyh}) are plainly consistent with their SDE description based on the PT-BFM approach and with the lattice ghost and gluon two-point Green's functions. We have also made fits with $a_{ln}^{(i)}$ fixed by the upper and lower bounds for the range of $c$ given in ref.~\cite{Athenodorou:2016oyh} and produced the red dotted lines in the plots. The yellow band in between depicts, precisely, the region where the SDE-based Ansatz fits optimally the data, in consistence with the lattice two-point Green's functions. 

\begin{table}
\begin{tabular}{||c||c||c||}
\hline 
\hline
& $i$=sym & $i$=asym \\
\cline{2-3}
\begin{tabular}{c}
\\ 
\hline
\hline
$a_0^{(i)}$ \\
\hline
$a_{ln}^{(i)}$ \\
\hline
$a_2^{(i)}$ \\
\hline 
$M$ [GeV]
\end{tabular}
&
 \begin{tabular}{c|c|c}
 Eq.(\ref{eq:gGamma0}) & Eq.(\ref{eq:gGammaF}) & Eq.(\ref{eq:gGammaF})
\\
  \hline
  \hline
 13.2-20.0 & 15.3-23.7& 20.6 \\ 
 \hline
2.05-3.12(*) & 2.05-3.12(*) & 2.74 \\
 \hline 
- & 21.3-41.7 & 34.4 \\
 \hline 
- & 0.78-0.71 & 0.72 \\
 \end{tabular}
 &  
 \begin{tabular}{c|c|c}
 Eq.(\ref{eq:gGamma0}) & Eq.(\ref{eq:gGammaF}) & Eq.(\ref{eq:gGammaF})
 \\
 \hline
 \hline
 12.4-18.8 &14.5-22.3 & 19.5 \\
 \hline
 1.92-2.91(*) & 1.92-2.91(*) & 2.55(*) \\
 \hline 
- &  26.1-42.5 &  36.6 \\
\hline 
- &  0.70-0.69 & 0.69 \\
 \end{tabular} 
 \\
\hline
\hline
\end{tabular}
\caption{\small Best-fit parameters obtained by applying the SDE-based Ansatz to describe the three-gluon 1-PI form factors in symmetric and asymmetric configurations. The first and fourth columns correspond to fits of \eq{eq:gGamma0} to, respectively, symmetric and asymmetric lattice data, with $a^i_{ln}$ fixed by the estimates (lower and upper bounds) of $c$ in ref.~\cite{Athenodorou:2016gsa}; the second and fifth stand for the fits with \eq{eq:gGammaF}, also with $a^i_{ln}$ fixed in the same way; and the third for a fit to the symmetric data with \eq{eq:gGammaF} and all the parameters free, while the sixth results from taking the value of $c$ derived from symmetric data and so fixing $a^{asym}_{ln}$ in a fit of \eq{eq:gGammaF} to the asymmetric data. The asterisk indicates that numbers do not result from a fit but are imposed, as above explained. The renormalization point is $\mu$=4.3 GeV. }
\label{Tab:param}
\end{table}

The position of the zero crossing depends on the competition of the logarithm coming from the ghost loop, controlled by the coefficient $c$, against the effect of the finite subleading corrections, the details of which we only precise here by a direct fit to three-gluon lattice data. Applying the best-fit of \eq{eq:gGammaF} to the lattice data, the zero crossing is found to lie on $p_0^{sym} \simeq 0.17$ GeV, for the symmetric case, and $p_0^{asym} \simeq 0.16$ GeV, for the asymmetric; while, had we employed \eq{eq:gGamma0} with the best-fit parameters of  Tab.~\ref{Tab:param}, one would obtain: $p^{i}_0 \simeq 0.17$ GeV, for both the symmetric and the asymmetric cases. Of course, there are non-negligible statistic and systematic uncertainties for these results. For instance, we estimate a relative error of around 15 \% for the position of the zero crossing, when using \eq{eq:gGamma0} in the symmetric case; while, in the much noisier asymmetric case, we get a relative error of 34 \%. Both are just statistical errors. In this work we do not aim on a very precise determination of the zero crossing but on a very strong confirmation that, after the appropriate projection, the nonperturbative ghost-loop contribution induces a logarithmic singularity at vanishing momentum for the three-gluon form factors, -independently of which particular momenta configuration we are considering-, which drives the vertex under question from positive to negative values.

\section{Conclusions}
\label{sec:conclusions}

We have investigated further on the IR structure of the three-gluon vertex, specially by studying the nonperturbative form factor associated to the tree-level tensor, precisely the one that should be invoked in the definition of the running strong coupling under the MOM renormalization prescription. Previous studies, both in lattice and continuous QCD --following various distinct approaches--, have provided with evidence for a change of sign of this form factor at a given momentum lying in the deep IR domain, at least for QCD without light quarks. It has been argued that the effect of light quarks would be only quantitative and consists of the shifting of the corresponding ``zero crossing" down to the IR. However, the issue has not been completely settled and one should be cautious about concluding the appearance of a zero-crossing in realistic QCD. 

In this work, still exploiting gauge-field configurations from quenched lattice QCD simulations but at very large physical volumes, we presented a more refined and elaborated analysis of the detection of the zero crossing for some three-gluon vertex form factors. This very specific infrared feature is posited to be closely related to the masslessness of the Faddeev-Popov ghost propagator circulating in the nonperturbative loop diagrams contributing to the gluon vacuum polarization, which is the underlying source of the negative singularity for the three-gluon form factors forcing the change of sign. Indeed, such a negative singularity is more striking and easily discernible as an effect of the nonperturbative ghost loop than its direct impact on the gluon propagator. Provided that the nonperturbative ghost loops have been also recently shown to have a noteworthy impact on the quark-gluon interaction kernel and, hence, on a process-independent effective strong coupling based on the PT-BFM approach~\cite{Binosi:2016nme}, the confirmation of the zero crossing and the numerical estimate of the ghost-loop impact (expressed by the coefficient $c$ in \eq{eq:Deltam1}) are also interesting results with possible phenomenological implications.

Compared to previous studies of the topic, we provided here more and stronger corroborative evidence to the statement in question. In particular by pushing the statistics and by employing different lattice actions (Wilson plaquette and tree-level Symanzik improved) we have under a much tighter control statistical and systematical uncertainties which lead us to make more clear statements both on the symmetric as well as the asymmetric momentum configuration. We provided a fully comprehensive and refined analysis grounded on its understanding within the framework of SDE under the PT-BFM scheme and recapitulated the basic argument on how, within this approach, the nonperturbative ghost-loop diagrams can lead to such a remarkable effect. On the basis of both our employment of pure gauge simulations with large physical volumes and our refined analysis, we managed not only to further solidify the zero crossing of the symmetric kinematical set-up but also of the much noisier asymmetric one that was previously smudged due to high statistical fluctuations.

\section*{Acknowledgements} 

We thank the support of Spanish MINECO FPA2014-53631-C2-2-P  research project, SZ acknowledges support by the National Science Foundation (USA) under grant PHY-1516509 and by the Jefferson Science Associates, LLC under U.S. DOE Contract $\#$ DE-AC05-06OR23177. We thank A. Athenodorou, D. Binosi and J. Papavassiliou for very fruitful discussions and for their participation in a previous work triggering this one. SZ is indebted to A. Sciarra for all his help regarding the CL2QCD code. CL2QCD is a Lattice QCD application based on OpenCL, applicable to CPUs and GPUs.
 Numerical computations have used resources of CINES and GENCI-IDRIS
as well as resources at the IN2P3 computing facility in Lyon.


\bibliography{total}

\end{document}